\documentclass[numberedappendix]{emulateapj}
\usepackage{apjfonts, natbib}

\usepackage{amsmath}
\usepackage{graphicx}

\newcommand{\be}{\begin{equation}}
\newcommand{\ee}{\end{equation}}
\def\be{\begin{equation}}
\def\ee{\end{equation}}
\def\ba{\begin{eqnarray}}
\def\ea{\end{eqnarray}}
\def\nn{\nonumber}
\def\lgl{\langle}
\def\rgl{\rangle}

\def\lsim{\mathrel{\rlap{\lower4pt\hbox{\hskip1pt$\sim$}}
    \raise1pt\hbox{$<$}}}               
\def\gsim{\mathrel{\rlap{\lower4pt\hbox{\hskip1pt$\sim$}}
    \raise1pt\hbox{$>$}}}    
\newcommand{\ld}{{\ell '}}
 
\newcommand{\lm}{{\ell m}}

\newcommand{\lld}{{\ell \ell'}}
\newcommand{\ldl}{{\ell' \ell}}
 
\def\mA{\mathbf{A}}
\def\mN{\mathbf{N}}
\def\vd{\mathbf{d}}
\def\vm{\mathbf{m}}
\def\vg{\mathbf{g}}

\shorttitle{Improved results from the QUaD CMB experiment}
\shortauthors{QUaD collaboration -- M.\,L.\,Brown et al.}

\begin{document}

\slugcomment{Submitted to ApJ}

\title{Improved measurements of the temperature and polarization of the CMB from QUaD}

\author{
  QUaD collaboration -- 
  M.\,L.\,Brown\altaffilmark{1,2},
  P.\,Ade\altaffilmark{3},
  J.\,Bock\altaffilmark{4,5},
  M.\,Bowden\altaffilmark{3,6},
  G.\,Cahill\altaffilmark{7},
  P.\,G.\,Castro\altaffilmark{8,9},
  S.\,Church\altaffilmark{6},
  T.\,Culverhouse\altaffilmark{10},
  R.\,B.\,Friedman\altaffilmark{10},
  K.\,Ganga\altaffilmark{11},
  W.\,K.\,Gear\altaffilmark{3},
  S.\,Gupta\altaffilmark{3},
  J.\,Hinderks\altaffilmark{6,12},
  J.\,Kovac\altaffilmark{5},
  A.\,E.\,Lange\altaffilmark{5},
  E.\,Leitch\altaffilmark{4,5},
  S.\,J.\,Melhuish\altaffilmark{13},
  Y.\,Memari\altaffilmark{8},
  J.\,A.\,Murphy\altaffilmark{7},
  A.\,Orlando\altaffilmark{3,5}
  C.\,O'\,Sullivan\altaffilmark{7},
  L.\,Piccirillo\altaffilmark{13},
  C.\,Pryke\altaffilmark{10},
  N.\,Rajguru\altaffilmark{3,14},
  B.\,Rusholme\altaffilmark{6,15},
  R.\,Schwarz\altaffilmark{10},
  A.\,N.\,Taylor\altaffilmark{8},
  K.\,L.\,Thompson\altaffilmark{6},
  A.\,H.\,Turner\altaffilmark{3},
  E.\,Y.\,S.\,Wu\altaffilmark{6}
  and
  M.\,Zemcov\altaffilmark{3,4,5} 
}

\altaffiltext{1}{Cavendish Astrophysics,
  University of Cambridge, J. J. Thomson Avenue, Cambridge CB3 OHE,
  UK.}
\altaffiltext{2}{Kavli Institute for Cosmology Cambridge, Madingley
  Road, Cambridge CB3 OHA, UK.}
\altaffiltext{3}{School of Physics and Astronomy, Cardiff University,
  Queen's Buildings, The Parade, Cardiff CF24 3AA, UK.}
\altaffiltext{4}{Jet Propulsion Laboratory, 4800 Oak Grove Dr.,
  Pasadena, CA 91109, USA.}
\altaffiltext{5}{California Institute of Technology, Pasadena, CA
  91125, USA.}
\altaffiltext{6}{Kavli Institute for Particle Astrophysics and
  Cosmology and Department of Physics, Stanford University,
  382 Via Pueblo Mall, Stanford, CA 94305, USA.}
\altaffiltext{7}{Department of Experimental Physics,
  National University of Ireland Maynooth, Maynooth, Co. Kildare,
  Ireland.}
\altaffiltext{8}{Institute for Astronomy, University of Edinburgh,
  Royal Observatory, Blackford Hill, Edinburgh EH9 3HJ, UK.}
\altaffiltext{9}{{\em Current address}: CENTRA, Departamento de F\'{\i}sica,
  Edif\'{\i}cio Ci\^{e}ncia, Piso 4, Instituto Superior T\'ecnico -
  IST, Universidade T\'ecnica de Lisboa, Av. Rovisco Pais 1, 1049-001
  Lisboa, Portugal.}
\altaffiltext{10}{Kavli Institute for Cosmological Physics,
  Department of Astronomy \& Astrophysics, Enrico Fermi Institute, 
  University of Chicago, 5640 South Ellis Avenue, Chicago, IL 60637, USA.}
\altaffiltext{11}{APC/Universit\'e Paris 7 -- Denis Diderot/CNRS,
  B\^atiment Condorcet, 10, rue Alice Domon et L\'eonie Duquet, 75205
  Paris Cedex 13, France.}
\altaffiltext{12}{{\em Current address}: NASA Goddard Space Flight
  Center, 8800 Greenbelt Road, Greenbelt, Maryland 20771, USA.}
\altaffiltext{13}{School of Physics and Astronomy, University of
  Manchester, Manchester M13 9PL, UK.}
\altaffiltext{14}{{\em Current address}: Department of Physics and Astronomy, University
  College London, Gower Street, London WC1E 6BT, UK.}
\altaffiltext{15}{{\em Current address}:
  Infrared Processing and Analysis Center,
  California Institute of Technology, Pasadena, CA 91125, USA.}

\begin{abstract}
We present an improved analysis of the final dataset from the QUaD
experiment. Using an improved technique to remove ground
contamination, we double the effective sky area and hence increase the
precision of our CMB power spectrum measurements by $\sim30\%$ versus
that previously reported. In addition, we have improved our modeling
of the instrument beams and have reduced our absolute calibration
uncertainty from 5\% to 3.5\% in temperature. The robustness of our
results is confirmed through extensive jackknife tests and by way of
the agreement we find between our two fully independent analysis
pipelines. For the standard 6-parameter $\Lambda$CDM model, the
addition of QUaD data marginally improves the constraints on a number
of cosmological parameters over those obtained from the WMAP
experiment alone. The impact of QUaD data is significantly greater for
a model extended to include either a running in the scalar spectral
index, or a possible tensor component, or both. Adding both the QUaD
data and the results from the ACBAR experiment, the uncertainty in the
spectral index running is reduced by $\sim25\%$ compared to WMAP
alone, while the upper limit on the tensor-to-scalar ratio is reduced
from $r < 0.48$ to $r < 0.33$ (95\% c.l). This is the strongest limit
on tensors to date from the CMB alone. We also use our polarization
measurements to place constraints on parity violating interactions to
the surface of last scattering, constraining the energy scale of
Lorentz violating interactions to $< 1.5 \times 10^{-43}$ GeV (68\%
c.l.). Finally, we place a robust upper limit on the strength of the
lensing $B$-mode signal. Assuming a single flat band power 
between $\ell = 200$ and $\ell = 2000$, we constrain the amplitude of 
$B$-modes to be $< 0.57 \, \mu {\rm K}^2$ (95\% c.l.).
\end{abstract}

\keywords{CMB, anisotropy, polarization, cosmology}

\section{Introduction}
\label{sec:intro}
\setcounter{footnote}{0} 
Observations of the polarization of the cosmic microwave background
(CMB) represent one of the most powerful probes available for
investigating the physics of the early universe (see
e.\,g. \citealt{challinor09} for a review). The CMB polarization field
can be decomposed into two independent modes: even parity $E$-modes are
generated, at the time of last scattering, by both scalar and tensor
(gravitational wave) metric perturbations. In contrast, odd parity
$B$-modes are generated at last scattering only by gravitational waves, a
generic prediction of inflation models. On small scales, $B$-modes
are also expected to arise from gravitational lensing of the $E$-mode
signal by intervening large-scale structures. A detection of $B$-mode
polarization (on any scale) has yet to be made. 

After the initial detection of the much stronger $E$-mode polarization
\citep{kovac02}, steady improvements have been made in measuring the
$E$-mode signal by a number of experiments \citep{leitch05, barkats05,
readhead04, montroy06, sievers07, page07, wu07, ade08, bischoff08,
nolta09}. Recently, a major step forward in precision CMB polarization
measurements was achieved with the high-significance detection of a
characteristic series of acoustic peaks in the E-mode polarization
power spectrum with our initial analysis of the final QUaD dataset
(\citealt{pryke09}; hereafter Paper II). For our analysis presented in
Paper II, in order to mitigate against a strong polarized ground
contaminant, we employed the technique of lead-trail
differencing. Although this technique is extremely successful, it does
have one major disadvantage --- the effective sky area is halved
(while the signal-to-noise is kept the same) resulting in a
corresponding increase of $\sim 40\%$ in the uncertainties on the
final power spectrum estimates. The major improvement which we implement in
this new analysis is a technique to remove the ground contamination
while preserving the full sky area. Our analysis thus yields
constraints on all six possible CMB power spectra which are
approximately $30\%$ stronger than those presented in Paper II.

We have also refined our modeling of the QUaD beams. For our previous
analysis, we modeled the beams as elliptical Gaussian functions. In
addition to the main lobe, there is a small sidelobe component ---
with our increased sensitivity, we now find it necessary to explicitly
model this sidelobe component. Accounting for the sidelobes results in 
a small ($\sim 10\%$) increase in the amplitude of our power spectrum
measurements on small scales (multipoles, $\ell \gsim 700$).

Note that in this paper, we present the results obtained from two
independent analysis pipelines. These are arbitrarily denoted
Pipeline 1 and Pipeline 2 and are derived from the two pipelines used
to analyze the data from our first year of observations
\citep{ade08}. The analysis presented in Paper II was
performed using Pipeline 2.

The paper is organized as follows. In Section~\ref{sec:obs_lowlevel},
we briefly summarize the QUaD observations and low-level processing
which are unchanged for this analysis. In
Section~\ref{sec:map_making}, we present our new technique for
mitigating against contaminating ground pick-up and describe the
details of our map making procedure. A description of our improved
beam modeling is given in Section~\ref{sec:beam_model} and our
treatment of the uncertainties is given in Appendix~\ref{app:beam_model_app}.
The absolute calibration of QUaD is briefly described in
Section~\ref{sec:abs_cal} with an error analysis given in 
Appendix~\ref{app:abs_cal_app}.  The results from our two independent
power spectrum analyses are presented in
Section~\ref{sec:power_spectra}. In Section~\ref{sec:parameters}, we
combine the QUaD results with data from the WMAP, ACBAR and SDSS
experiments to place constraints on the parameters of a number of
cosmological models. Our conclusions are presented in
Section~\ref{sec:conclusions}.

\section{Observations and low-level processing}
\label{sec:obs_lowlevel}
The QUaD experiment and its performance are described in
\cite{hinderks09}, hereafter referred to as Paper I. The low-level data
processing is described in Paper II. The initial low-level processing
of the raw data has not changed for the analysis presented in this
paper so here we give only a brief summary and refer the reader to
Paper II for a detailed description.

The QUaD experiment was a 2.6m Cassegrain radio telescope which
observed from the South Pole for three seasons from 2005 to 2007. The
QUaD receiver consisted of 31 pairs of (orthogonal) polarization
sensitive bolometers (PSBs), 12 at 100~GHz and 19 at 150~GHz. These
PSB pairs were arranged on the focal plane in two orientation angle
groups separated by 45$^\circ$. The raw time-ordered data (TOD), which
were sampled at 100 Hz, were first deconvolved to correct for the
finite response times of the PSBs and electronics used to detect the
incoming signal. The time-constants used for each detector were
measured using an external Gunn oscillator source as described in
Paper I. After deconvolution, the detector data were low-pass filtered
to $< 5$~Hz.\footnote{%
QUaD's CMB observations employed a relatively slow scan speed of
$0.25^\circ$/~sec. For our observing declination ($\sim -50^\circ$),
the sky signal for multipoles, $\ell < 2000$, appears in the
time-stream at $< 1$~Hz. The low-pass filtering therefore removes none
of the sky signal (for $\ell < 2000$) but it does remove high
frequency noise introduced by the deconvolution procedure.}  
The data were then de-glitched to remove cosmic rays and other events.
A relative calibration was then applied to each detector using the
"elevation nod" technique described in Paper I and Paper II.

Once de-glitched and calibrated, the data were downsampled to 20~Hz.
For the analysis presented in this paper, we have retained the exact
same data cuts for bad weather, moon contamination and badly behaved
detectors as were used in Paper II. Out of a total of 289 days of
observations during 2006 and 2007, after applying these data-cuts, 143
remained for the science analysis. Although fully code independent,
the low level parts of our two analysis pipelines are algorithmically
similar.

\section{Map-making using ground template removal}
\label{sec:map_making}
Our improved technique for removing the ground signal relies on
redundancies in the scan strategy so before describing our templating
procedure, we turn first to the QUaD scan strategy and
examining the redundancies present within it.

\subsection{QUaD observing strategy}
\label{sec:obs_strategy}
QUaD observed a $\sim\!100$ square degree area of sky, centered on
RA 5.5h, Dec $-50^{\circ}$. The field is fully contained within the
shallow field observed by the 2003 flight of the Boomerang experiment
(hereafter referred to as B03, \citealt{masi06}). The QUaD field also
partially overlaps with B03's deep field. The QUaD
observations employed a lead-trail scheme, whereby each hour of
observations were split equally between two adjoining subfields,
separated in RA by 0.5 h --- the lead field, centered on RA 5.25 h, and
the trail field, centered on RA 5.75 h.

The scanning strategy consisted of constant-elevation scans back and
forth over a 7.5 deg throw in azimuth, applied as a
modulation on top of sidereal tracking of the field
center. Each hour of observation was equally split between the lead
and trail fields. These half-hour sessions were further divided into
four "scan-sets", consisting of ten "half-scans" each, and the telescope
was stepped in elevation by 0.02 degs between scan-sets. After a half
hour scanning the lead field, the telescope pointing returned to
its starting position in azimuth and elevation, and repeated the same scan pattern
with respect to the ground, but now scanning the trail CMB
field. The trail field's scan pattern was thus a replica (in
azimuth/elevation coordinates) of the lead field's.
After an hour the pointing moved on to a fresh part of sky and the
process repeated. This scan pattern was designed to facilitate the
lead-trail differencing analysis presented in Paper II, whereby each
pair of lead-trail partner scans are point by point
differenced. Any ground signal, which is stable in time over the half
hour which separates the lead and trail observations, will be
completely removed by this differencing, at the expense of a reduction
in the effective sky area by a factor of two.\footnote{%
For a Gaussian field, such as the CMB, neglecting correlations between
the signal in the lead and trail fields, the fluctuations in the
differenced field will be amplified by a factor of $\sqrt{2}$ and the power
spectrum will increase by a factor of 2. The noise is amplified in a
similar manner and so the signal-to-noise ratio in the differenced
field remains unchanged from that achieved in the non-differenced
field.}

In addition to the lead-trail scheme, a further redundancy is present
in the scan-strategy due to the movement of the CMB field across the
sky during each scan-set. The time elapsed from when a given sky pixel
is first visited on the first half scan of a set, to when it is last
visited on the tenth half-scan is $\sim 6$ minutes. During this
time the sky rotates by 1.5 degrees. Using only data from a single
scan-set, there is therefore scope for separating signals originating
on the ground from those originating on the sky, on scales smaller than
1.5 degrees, corresponding to $\ell \sim 250$ in multipole space. One
can achieve further separation of ground from sky by combining the
data from lead and trail partner observations. For this work we have
made use of both the lead-trail and sky rotation redundancies to
separate the ground and sky signals, and to reconstruct the CMB fields
over the full sky area.

\subsection{Ground template removal}
\label{sec:template_removal}
Field differencing is a sub-optimal use of the
redundancy in the scan strategy to mitigate against ground pickup. 
One can retain more of the sky information by constructing and removing
estimates of the ground signal. To facilitate the removal of ground
signal, we can model the TOD as
\be
d_i = S_i(\theta) + g(\alpha) + n_i + o_{\rm scan} \, ,
\label{eqn:tod_model1}
\ee
where $S_i(\theta)$ is the sky signal,
\be
S_i(\theta) = \frac{1}{2} \left[I (\theta) + Q(\theta)\cos(2\phi_i)
  +U(\theta)\sin(2\phi_i)\right] \, . 
\label{eqn:tod_model2}
\ee
Here, $I, Q$ and $U$ are the Stokes parameters in the direction,
$\theta$ on the sky and $\phi_i$ is the polarization sensitivity angle
(a combination of detector orientation on the focal plane and
boresight rotation) of each detector. 
 
In equation~(\ref{eqn:tod_model1}), $g(\alpha)$ represents the ground
signal as a function of azimuth, $\alpha$. In what follows, we will be
constructing and removing estimates of $g(\alpha)$ for each QUaD
detector and for each pair of lead and trail scan-sets
independently. Within these subsets of the data, the elevation is
constant and so the ground signal estimates (which we refer to as
ground ``templates'') are constructed as a function of azimuth
only. However, since we allow the templates to differ between
detectors and between lead-trail scan-set pairs, in practice, the
ground signal which we remove from the data does depend on elevation
also. Moreover, since templates are constructed for each detector
individually, our model also allows the ground signal to depend on
frequency. 

The resolution with which to construct the ground templates, in
general, needs to be determined through trial and error. In practice,
we find that the effect of changes in this parameter on the resulting
maps are imperceptible. This suggests that the ground signal varies
smoothly in azimuth and that a fairly coarse resolution is sufficient
to characterize it. For this analysis, we have used a resolution of
$\Delta \alpha = 0.1$ degs, midway between the minimum and maximum
resolutions we have investigated.
 
In equation~(\ref{eqn:tod_model1}), we have split the noise component
into a random part ($n_i$, which we model as a Gaussian random
variable; see Section~\ref{sec:sims}) and an offset ($o_{\rm scan}$)
which we model as a constant for each half-scan. We have found that it
is essential to remove these offsets before constructing ground
templates from the data. Otherwise the resulting templates tend to be
dominated by the long-timescale part of the $1/f$ atmospheric noise
(i.e. \!the offsets) rather than the ground signal which we are
attempting to characterize.

Regardless of the technique employed to reconstruct maps of the Stokes
parameters (naive, maximum-likelihood etc.), one can re-cast the well
known map-making equation (e.g. \citealt{stompor02} and references therein),
\be
\mA^T \mN^{-1}\mA \vm = \mA^T \mN^{-1} \vd \, , 
\label{eqn:mapmake}
\ee
to reconstruct both the sky signal and the ground signal simply by
making the substitutions,
\ba
\vm  &\rightarrow& \vm + \vg \nonumber \\
\mA  &\rightarrow& (\mA^{CMB}, \mA^G) \, , 
\ea
where $\vm$ is the reconstructed CMB map, $\vg$ is the reconstructed
ground signal and $\mA^{CMB}$ and $\mA^G$ are the "pointing matrices"
associated with the CMB and ground signals respectively. For a highly
redundant scan strategy, equation~(\ref{eqn:mapmake}) should be
soluble exactly and the CMB and ground signals should be completely
separable. However, for a scan strategy such as that used for QUaD
with limited revisiting of the same sky pixels at
different azimuths, this complete separation between sky and ground is
not possible. In this case, one can still separate the ground and sky
signals on smaller scales but one loses all information on the
largest scales where the separation is degenerate. For QUaD, we find
that below $\ell \sim 200$ our template removal procedure offers
essentially no improvement over field-differencing and that the
separation is near perfect by $\ell \sim 1000$.

Attempting to simultaneously solve for the sky and ground signals in
the QUaD data by applying equation~(\ref{eqn:mapmake}), we find that large scale
(ground signal) gradients are introduced to the resulting CMB maps due
to the degeneracy between the CMB and the ground signal on the largest
scales. One could certainly modify the procedure (e.g. by
marginalizing over the large-scale CMB and ground-signal modes) to
solve this problem. For the analysis presented here, we adopt a
simpler approach and simply solve for the ground signal
independently, subtract this from the TOD, and then construct the CMB
maps from the ground-cleaned TOD. We account for the
resulting filtering of the CMB signal in our Monte-Carlo
analysis. Our analysis assumes that the ground signal does not change between the start of
the lead scan set and the end of the corresponding trail scan set ($\sim 36$
mins). This is only a slight relaxation of the assumption that was
made for our previous analysis where we assumed that the ground signal
was constant over a 30 minute timescale. 

To apply the template removal, we proceed as follows. First, we
estimate and remove the atmospheric $1/f$ offsets, $o_{\rm scan}$ from
each half scan. To estimate the offsets, we simply take the mean of
the data within each scan. Note however that we restrict the azimuth
range over which we calculate the offsets to the central azimuth range
where all the scans within a scan-set overlap. This ensures
that our estimated ground templates will be unbiased over the full
azimuth range.

After removing the offsets, templates of the ground signal are
constructed by simple binning of the timestream data in azimuth. That
is, for each ground template ``pixel'', we construct
\be
\hat{g}(\alpha) = \frac{1}{N_{\rm hits}} \sum_{i \in \Delta \alpha}
d_i \, ,
\label{eqn:template_estimates}
\ee
where the sum is over all data from the lead scan set and its partner
trail scan set which falls in the azimuth range, $\Delta \alpha$. 

\begin{figure*}[t]
  \vspace{0.3cm}
  \centering
  \resizebox{1.0\textwidth}{!}{  
    {\includegraphics{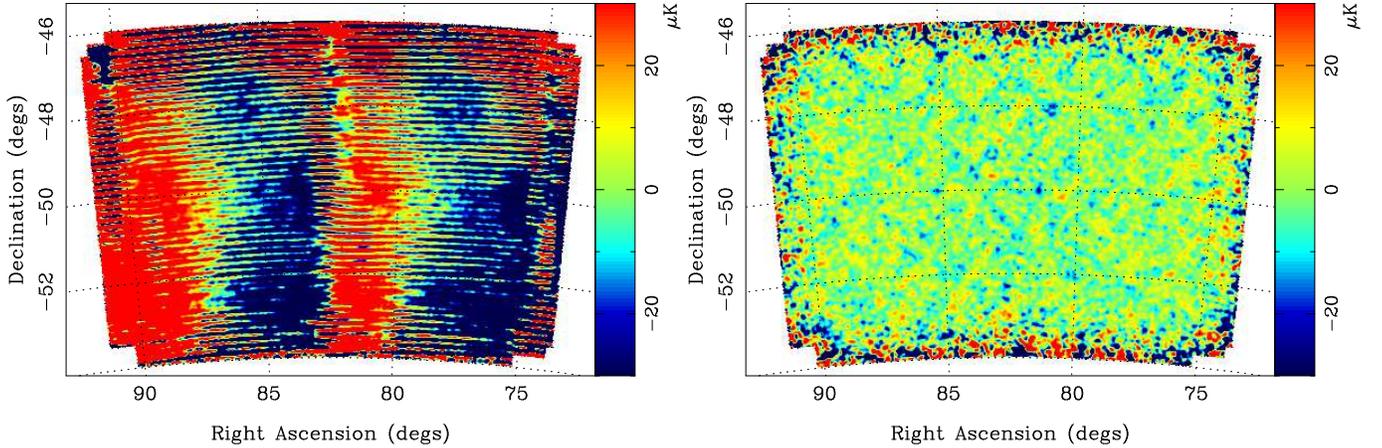}}}
  \vspace{0.2cm}
  \caption{Demonstration of the performance of the ground templating
    procedure described in the text. The figure shows maps of the
    Stokes $U$ polarization over the full QUaD sky area at 150~GHz,
    smoothed with a 5 arcmin Gaussian kernel. The map on the left is
    that obtained without removing ground templates and is heavily
    contaminated by ground pickup. Note the similarity of the
    contamination between the lead and trail halves of the map. The
    map on the right is that obtained when we include our templating
    procedure. Clearly the vast majority of the ground signal is
    successfully removed by this process. For the purposes of this
    illustration, in order to highlight the success of the template
    removal, in both cases, only the mean of each scan was removed
    from the TOD before mapping. However, for our cosmological analysis, we use
    maps which have had third-order polynomials removed from each scan
    (see Figure~\ref{fig:tqu_maps}).}
  \label{fig:umaps_smooth}
  \vspace{0.5cm}
\end{figure*}

Although they will be unbiased estimates of the ground signal, the
templates constructed using equation~(\ref{eqn:template_estimates})
will also contain CMB signal and noise. The expectation value of
the constructed templates is
\be
\lgl \hat{g}(\alpha) \rgl = g(\alpha) + \frac{1}{N_{\rm hits}} \left[ \sum_{i \in
\Delta \alpha} S_i(\theta) + \sum_{i \in \Delta \alpha} n_i \right] \, ,
\label{eqn:template_expvals}
\ee 
where $S_i(\theta), g(\alpha)$ and $n_i$ are the quantities defined in
equation~(\ref{eqn:tod_model1}). In the limit of a highly redundant scan
strategy and uncorrelated noise, the terms in brackets will average to
zero and the templates will contain only ground signal.\footnote{%
For an experiment sensitive to absolute temperature, the
first term in brackets in equation~(\ref{eqn:template_expvals}) would,
in fact, average to the CMB monopole for a highly redundant scan
strategy. For an experiment sensitive to temperature differences only
(such as QUaD), this term would average to zero.}
For a realistic scan strategy and correlated noise, these terms will be
non-zero and so removing the templates will have the effect of
filtering both large-scale CMB signal and long-timescale noise. We
correct for this filtering by including the template removal procedure
in our Monte-Carlo simulations (Section~\ref{sec:sims}).

Finally, to obtain estimates of the TOD which are free of ground
pickup, the templates are subtracted from the original TOD: 
\be
\vd^{\rm clean} = \vd - \mA^G \hat{\vg} \, ,
\label{eqn:template_subtract}
\ee
where $\mA^G$ is the pointing matrix associated with the ground signal
and $\hat{\vg}$ is the estimated ground signal constructed using
equation~(\ref{eqn:template_estimates}). This
will result in TOD which is, in principle, free of ground contamination
and can thus be modeled as in equation~(\ref{eqn:tod_model1}) but now
without the $g(\alpha)$ term. Once the ground
contamination has been removed, our map-making proceeds as
described in Paper II. Explicitly, we perform the following operations. For
each azimuth scan the best-fit third order polynomial is subtracted 
to remove the long timescale part of the atmospheric $1/f$
noise. For each PSB pair, the data is then summed and
differenced to yield temperature ($s_i$) and polarization ($d_i$)
TOD. Maps of the temperature (or Stokes $I$) CMB field are then
constructed from the summed data using a simple weighted average:
\be
T = \frac{1}{\sum_i w_i} \sum_i w_i s_i \, ,
\label{eqn:tod_weights}
\ee
where the weights are given by $w_i = W(x)/v_{\rm scan}$. Here, $v_{\rm
scan}$ is the variance of the data across the parent half-scan ---
noisy data (e.g. \!due to bad weather) is thus down-weighted. $W(x)$, where $x$ denotes the
fractional position within the scan, is an apodization (the same for
each scan) which we use to down-weight the scan ends. We apply this
apodization to reduce the tiling effects seen in our previous analysis
(see Section 6.3 and Figure 15 in Paper II) whereby the interaction of
the polynomial filtering with different sky coverage for different
detectors produced visible step features in the final maps. 
The exact form used for the apodization is not important. 

We construct maps of the Stokes polarization parameters, $Q$ and $U$ as
\ba
\left( \begin{array}{c} Q \\ U \end{array} \right) &=& \left(
  \begin{array}{cc} 
    \lgl\cos^2(2\phi_i)\rgl & \lgl\cos(2\phi_i)\sin(2\phi_i)\rgl \\
    \lgl\cos(2\phi_i)\sin(2\phi_i)\rgl & \lgl\sin^2(2\phi_i)\rgl \\ 
\end{array} \right)^{-1} \nn \\
&&\mbox{} \times \left( \begin{array}{c}
    \lgl\cos(2\phi_i)d_i\rgl \\
    \lgl\sin(2\phi_i)d_i\rgl \\ \end{array} \right) \, ,
\label{eqn:qu_mapmaking}
\ea
where the angled brackets denote an average taken over all data
falling within each map pixel and the angle, $\phi_i$ is a combination
of the polarization sensitivity direction of each detector on the
focal plane and the "deck angle" (rotation about the telescope
boresight) of the observation. Note that to construct the averages
(e.g. $\lgl\cos(2\phi_i)d_i\rgl$) on the right-hand side of
equation~(\ref{eqn:qu_mapmaking}), we also use inverse-variance
weights as in equation~(\ref{eqn:tod_weights}).

Note that the only difference between the approaches of our two
pipelines in applying the above operations is during the final
coaddition of the template-subtracted data into CMB maps: Pipeline 1
uses HEALPix\footnote{%
See http://healpix.jpl.nasa.gov/index.shtml and \cite{gorski05}.}
to pixelize the sky at a resolution of $\sim\!1.7$ arcmin
($N_{\rm side} = 2048$). Pipeline 2 works under the flat sky
approximation and pixelizes the sky into a 2D cartesian grid with a
spacing of $1.2$ arcmin. Figure~\ref{fig:umaps_smooth} shows an
example of the performance of the template removal procedure (for
Pipeline 1) for the $150$~GHz Stokes $U$ polarization map. Note that,
in order to highlight the success of the template removal, for this
demonstration we have not applied the third order polynomial removal
mentioned above to the TOD and have only removed the mean from each
scan.\footnote{%
In fact, the polynomial fitting procedure does remove the gross
features of the ground signal from the data although much remains ---
full field maps which have not been subjected to template removal fail
jackknife tests at high significance regardless of whether we apply
polynomial removal or not.}

\begin{figure*}[t]
  \vspace{1.0cm}
  \centering
  \resizebox{1.0\textwidth}{!}{  
    {\includegraphics{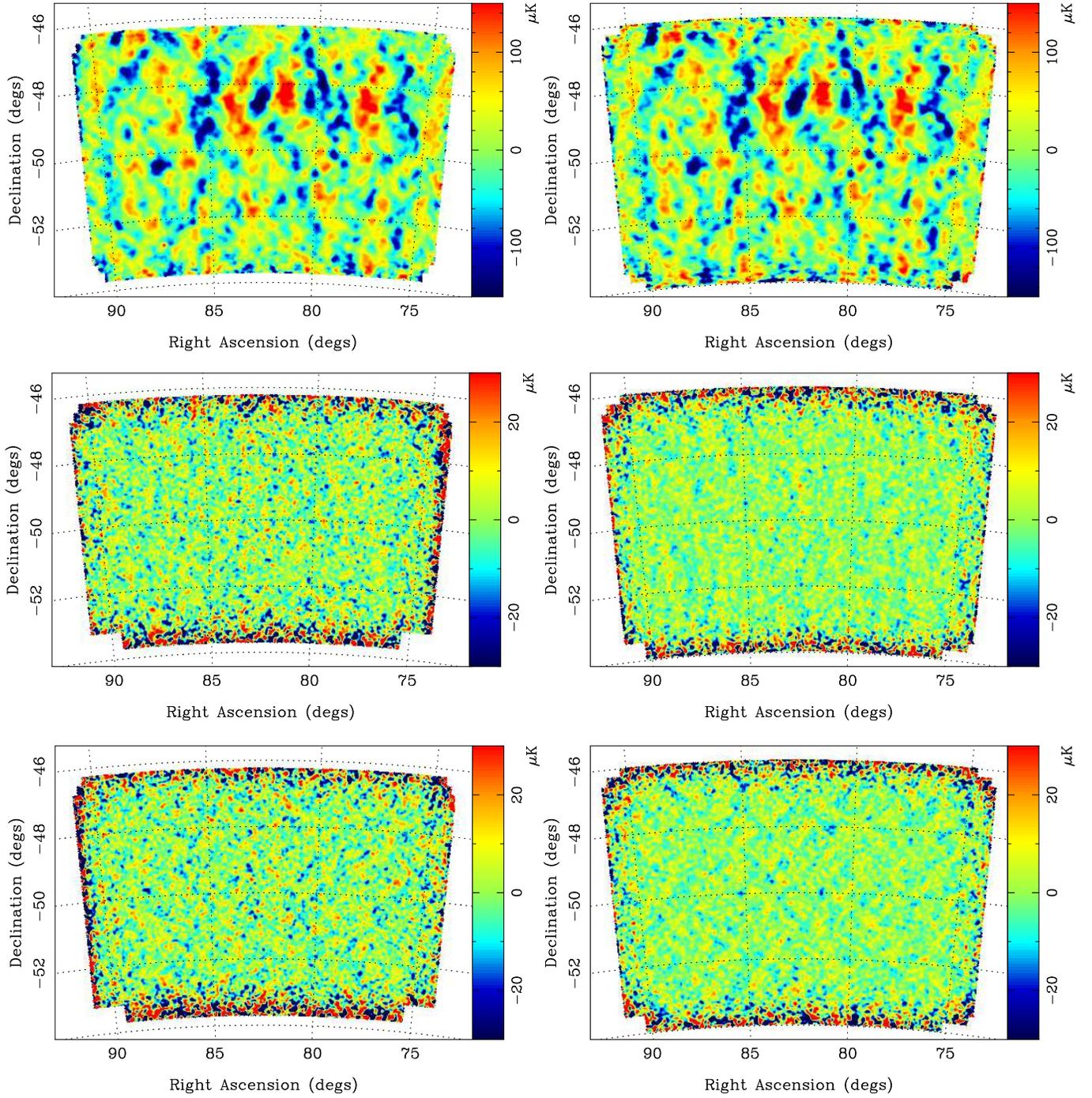}}}
  \vspace{0.2cm}
  \caption{Maps of the $T$ (top panels), $Q$ (middle panels) and $U$ (lower panels)
    Stokes parameters over the full QUaD sky area at 100~GHz (left) and
    150~GHz (right). For display purposes only, the maps have been
    smoothed with a 5 arcmin Gaussian kernel. Note the difference in
    the color stretch used to display the temperature and polarization
    maps.}
  \label{fig:tqu_maps}
\end{figure*}

In Figure~\ref{fig:tqu_maps}, we present the full set of maps
($T , Q$ and $U$ at $100$ and $150$~GHz; again for Pipeline 1) over
the full sky area as estimated using the template removal
procedure (now including the third order polynomial removal). For the
purposes of visual illustration only, we have smoothed
each of the maps with a 5 arcmin Gaussian kernel in order to bring out
the CMB structure.

\begin{figure*}[t]
  \vspace{0.3cm}
  \centering
  \resizebox{1.0\textwidth}{!}{  
    {\includegraphics{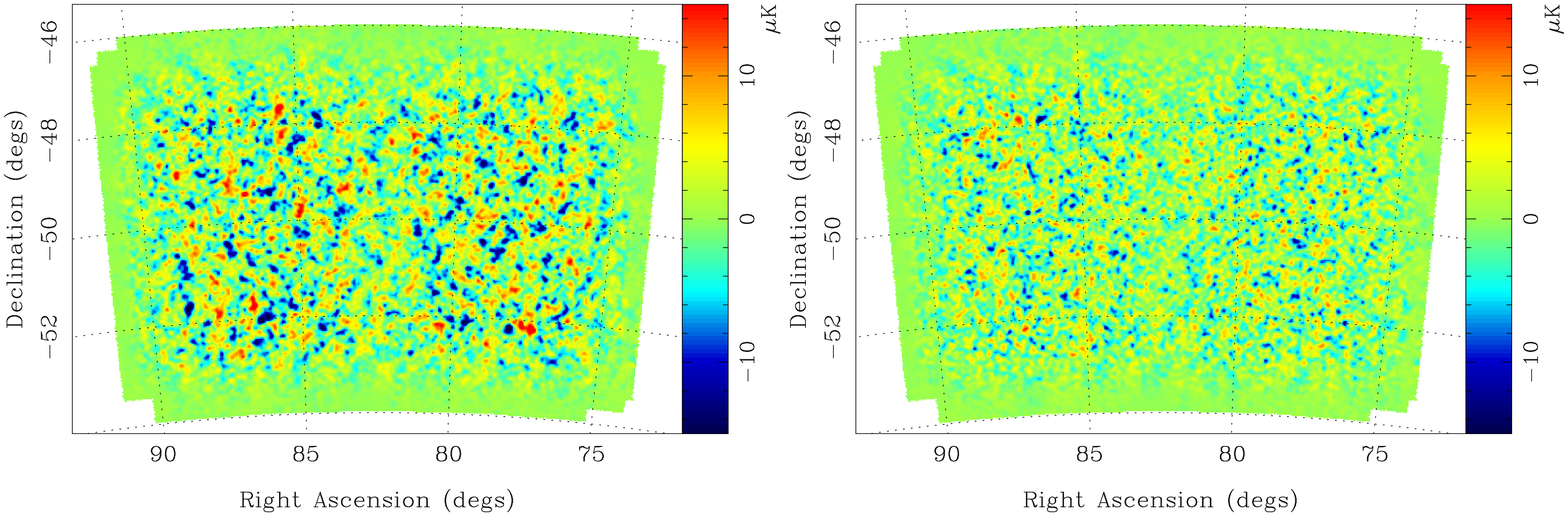}}}
  \vspace{0.2cm}
  \caption{Apodized 150~GHz QUaD polarization maps decomposed into
    $E$-modes (left) and $B$-modes (right). Once again, the maps have
    been smoothed with a 5 arcmin Gaussian kernel. The data is clearly
    dominated by $E$-modes. The amplitude of fluctuations present in
    the $B$-mode map is consistent with noise. The slight reduction in
    the amplitude of fluctuations towards the central RA of the field
    is due to the application of the apodization mask which
    down-weights the ``seam'' between the lead and trail halves of the
    map.}
  \label{fig:eb_maps}
  \vspace{0.5cm}
\end{figure*}

We can quickly (and crudely) assess the relative amounts of $E$- and
$B$-mode power in the polarization maps by decomposing the $Q$ and $U$
maps into $E$ and $B$ modes. We do this under the flat-sky
approximation. To minimize the impact of the noisy edge regions of the
maps, and to reduce the effects of $E/B$ mixing due to the finite
survey geometry, we apply an apodization to our maps before Fourier
transforming. ($E/B$ mixing is fully accounted for during our power
spectrum estimation described in Section~\ref{sec:power_spectra}.)
The resulting $E$ and $B$ maps at $150$~GHz, are shown in
Figure~\ref{fig:eb_maps}. We clearly detect significantly more
$E$-mode than $B$-mode structure.  The reconstructed $B$-mode map
shows similar levels of fluctuations to our polarization jackknife
maps (see Section~\ref{sec:jackknives}) and is consistent with noise.

\section{Beam measurements and modeling}
\label{sec:beam_model}
Our main set of observations for investigating the QUaD beam shapes
are a series of single day observations of the QSO PKS0537-441. 
In Paper II, our beam model consisted of elliptical Gaussian fits to
these QSO observations for each channel.

Given our increased sensitivity, we now include an additional sidelobe
component in our beam model. In order to measure the sidelobes from
the QSO data, we apply a 6th-order polynomial filter to the TOD before
mapping (with the QSO masked) and coadd these data over all channels
and over three days of observations. The radially averaged beam
profiles measured from these maps reveal the presence of sidelobe
structure at just below the -20 dB level, as predicted by the physical
optics (PO) simulations of QUaD presented in \cite{osullivan08}. Our two
analysis pipelines model these observations in slightly different ways
though both are matched to the QSO data.

Pipeline 1 rescales the PO models. The beam profiles which directly
result from the PO simulations are not a perfect match to the QSO
observations. In particular, the predicted main lobe widths are
smaller than observed while the predicted levels of sidelobes are
somewhat larger than observed. To match the PO models to the QSO data,
we parametrize the models using two parameters, one which scales the
main lobe width and one which varies the amplitude of the
sidelobes. We then use the observed QSO radial profiles to fit for
these parameters. The resulting best-fit re-scaled PO models are used
to model the beam.

Pipeline 2 models the beams in a fully empirical manner and is an
extension of the model used in Paper II. Using the existing elliptical
Gaussian fits to the quasar data, a pure Gaussian simulated beam
coadded across channels and observation dates is generated and
subtracted from the measured QSO maps. The residual after subtraction
is the sidelobe component of the beams. This residual is too noisy to
be used directly and so it is radially averaged to produce an
(assumed) azimuthally symmetric sidelobe template. This sidelobe
template is then added to the original Gaussian elliptical models to
produce a fully empirical beam model.

\begin{figure}[t]
  \vspace{0.3cm}
  \centering
  \resizebox{0.48\textwidth}{!}{  
    \rotatebox{-90}{\includegraphics{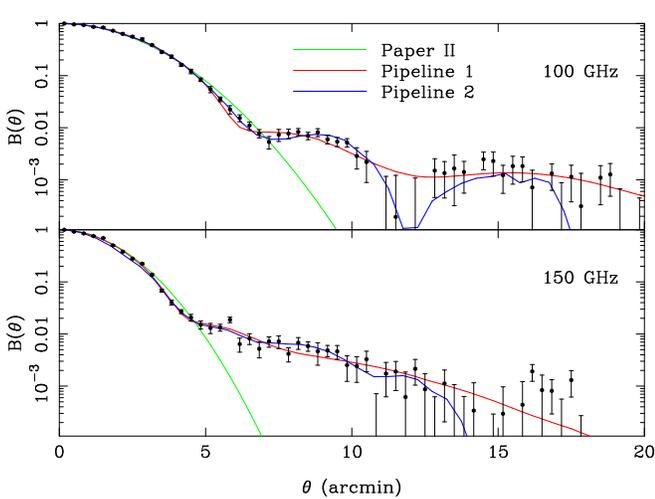}}}
  \caption{QUaD beam profiles at 100~GHz (top panel) and 150~GHz
  (bottom panel) as measured from the QSO PKS0537-441. The
  radial profiles as predicted using our new beam models for both
  pipelines are over-plotted as the red and blue curves and show good
  agreement with the QSO data. Also shown for comparison is the
  elliptical Gaussian beam model used for our previous analysis in
  Paper II.}
  \label{fig:beam_sidelobes}
\end{figure}

Figure~\ref{fig:beam_sidelobes} shows the radially averaged profiles
measured from the QSO data along with the profiles as predicted using
our old elliptical Gaussian beam model and as predicted using our
current beam models. Our revised beam models are clearly a superior
description of the true beams and are in good agreement --- in terms of
the resulting beam transfer functions, the two beam models agree to within
$4\%$ at 100~GHz and to within $2\%$ at 150~GHz for $\ell < 2000$.
A description of how we account for the remaining uncertainties on
our beams is given in Appendix~\ref{app:beam_model_app}.

\section{Absolute calibration}
\label{sec:abs_cal}
We derive the absolute calibration of QUaD by cross-correlating our
temperature maps with maps from B03 \citep{masi06}. This analysis is
done in spherical harmonic ($a_\lm$) space following the calibration
technique used by Boomerang which, in turn, was calibrated against the
WMAP 1st-year maps \citep{bennett03}.\footnote{%
We have also performed the calibration against the WMAP maps directly
and we find consistent results. However, calibrating using the B03
maps produces a more accurate result due to the larger overlap in
angular scale of B03 and QUaD.}  
We apply a correction to the B03 maps to account for the change in
calibration (1.25\% in temperature) between the 1- and 5-year WMAP
analyses \citep{hinshaw09}.  The B03 maps (which are essentially
unfiltered) are first passed through the QUaD simulation pipelines to
ensure that they are filtered in an identical manner to the QUaD
maps. Taking the spherical harmonic transforms of the maps, the
absolute calibration factors for QUaD are then given by

\be
S_\ell^\mathrm{quad} = \frac{ B_\ell^\mathrm{\, quad} \lgl
  a_\lm^\mathrm{\, b03 - X} a_\lm^{\ast \, \, \mathrm{b03 - Y}} \rgl}
                         { B_\ell^\mathrm{\, b03} \lgl a_\lm^\mathrm{\, b03
			     - X} a_\lm^{\ast \, \, \mathrm{quad}} \rgl} \, ,
\label{eqn:abscal}
\ee 
where the superscripts, $\!\mathrm{X}$ and $\!\mathrm{Y}$ denote two
noise-independent 145~GHz B03 maps and $B_\ell^\mathrm{\, quad}$ and
$B_\ell^\mathrm{\, b03}$ are the beam transfer functions of the two
experiments. This process produces an absolute calibration factor (in
$\mu$K/V) as a function of multipole, $\ell$. We take our final
absolute calibration factors to be the mean of this value in the range
$200 < \ell < 800$, corresponding to the overlapping scale range of
the two experiments. The largest contributors to our calibration error
are the quoted B03 uncertainty of $2\%$ and a relative pointing
uncertainty between QUaD and B03. Our final calibration uncertainty is
3.4\%. Further details on how we estimate this uncertainty are given
in Appendix~\ref{app:abs_cal_app}.
  
\section{Power spectrum analysis}
\label{sec:power_spectra}
To estimate the CMB power spectra, as in Paper II, we adopt a
Monte-Carlo (MC) based technique whereby we rely on accurate
simulations of the experiment to correct for the effects of noise,
beams, timestream filtering and the removal of the ground
templates. Before describing the MC simulations, we first describe the
differences between our two pipelines in their approach to power
spectrum estimation.

\subsection{Power spectrum estimation}
\label{sec:spectra_estimation}
Both of our pipelines broadly follow the so-called pseudo-$C_\ell$
technique \citep{hivon02}, extended to polarization
\citep{brown05}. Note that, for both pipelines, in order to minimize edge
effects, the $T, Q$ and $U$ maps are first apodized with an
inverse-variance mask as described in Section~6.1 of Paper II.

Pipeline 1 works on the curved sky and uses fast spherical harmonic
transforms to estimate the pseudo-$C_\ell$ spectra. These spectra are
then corrected for the effects of the sky cut, noise and filtering,
and binned into band powers according to
\be
{\bf P}_b=\sum_{b'}{\bf K}^{-1}_{bb'} \, \sum_\ell P_{b'\ell} \, 
(\widetilde{\bf C}_\ell-\lgl \widetilde{\bf N}_\ell \rgl_{MC}) \, .
\label{eqn:band_powers2}
\ee
In this equation and throughout this section, we use bold-face to
denote six-spectrum quantities, e.g.\! ${\bf P}_b$ is a $
6 \, n_{\, \rm band}$ vector, 
${\bf P}_b = \{P_b^{TT}, P_b^{EE}, P_b^{BB}, P_b^{TE}, P_b^{TB}, P_b^{EB}\}$, 
and ${\bf K}_{bb'}$ is a $6 \, n_{\, \rm band} \times 6 \, n_{\, \rm band}$
matrix, given by 
\be
{\bf K}_{bb'} = \sum_\ell P_{b \ell} \sum_\ld {\bf M}_{\lld} F_\ld
Q_{\ld b'} \, .
\ee
In equation~(\ref{eqn:band_powers2}), $\widetilde{\bf C}_\ell$ are the pseudo-$C_\ell$ spectra
estimated from the data maps and $\lgl \widetilde{\bf N}_\ell
\rgl_{MC}$ are the noise power spectra as measured from
simulations. 

$P_{b \ell}$ is a binning operator which bins the raw
pseudo-$C_\ell$ into band powers and $Q_{\ell b}$ is the inverse
operator which ``unfolds'' a band power into individual $C_\ell$s. For
this analysis, we use ``flat'' band powers for which the quantity,
$\ell(\ell + 1) C_\ell / 2 \pi$ is constant within each band. That is,
the binning operator we use is 
\be
P_{b \ell} = 
\left\{ \begin{array}{ll} 
\frac{1}{2 \pi} \frac{\ell(\ell+1)}{\ell^{(b+1)}_{\rm low} - 
\ell^{(b)}_{\rm low}}, & \mbox{  if  } 2 \le \ell^{(b)}_{\rm low} 
\le \ell < \ell^{(b+1)}_{\rm low} \\
0, & \mbox{  otherwise,  } \end{array} \right.
\ee
with the corresponding inverse operator given by
\be
Q_{\ell b}=
\left\{ \begin{array}{ll}
\frac{2 \pi}{\ell(\ell+1)}, & \mbox{  if  } 2 \le \ell^{(b)}_{\rm low} 
\le \ell < \ell^{(b+1)}_{\rm low} \\
0, & \mbox{  otherwise,  } \end{array} \right. 
\ee
where $\ell^{(b)}_{\rm low}$ denotes the nominal lower edge of band $b$.

The coupling matrix, ${\bf M}_{\lld}$, describes the mode-mixing effects of
the apodization mask and sky cut and is given in \cite{brown05} for
the full set of six possible CMB spectra. Note that the ${\bf
M}_{\lld}$ matrix fully encodes $E/B$ leakage effects due to the
finite survey geometry and so our Pipeline 1 estimator explicitly
corrects the $EE$ and $BB$ spectra for this leakage in the mean.

$F_\ell$ is a transfer function which we use to describe the combined
effects of timestream filtering, beam suppression and filtering of the
sky signal due to the removal of the ground templates. This function
will also encode any other signal suppression effects which are
present in our simulations (e.g. pixelization effects). We estimate
$F_\ell$ from our signal-only simulations as described in
Section~\ref{sec:transfer}.

Pipeline 2 works in the flat sky approximation and uses 2D FFTs to
estimate power spectra. This pipeline is described in detail in Paper
II. The power spectrum estimator for Pipeline 2 can effectively be
written as
\be
{\bf P}_b= \, F_b^{-1} \sum_\ell P_{b \ell} \, 
(\widetilde{\bf C}_\ell-\lgl \widetilde{\bf N}_\ell \rgl_{MC}) \, ,
\label{eqn:band_powers3}
\ee 
where $F_b$ is the binned equivalent of the per-multipole transfer
function, $F_\ell$ and we have implicitly made the connection between
the flat sky and curved sky power spectra, $C_\ell \approx P(k)$ for
$\ell \approx 2\pi k$. 

Note that (in addition to the flat sky approximation), the primary
difference between the two pipelines is that Pipeline 1 performs the
correction for the mode-mixing effects induced by the sky cut whereas
Pipeline 2 does not perform this correction. Because of this
difference, neither the recovered band powers nor their uncertainties
are directly comparable between the two pipelines. A proper comparison
of the two analyses requires the use of the associated band power
window functions (Section~\ref{sec:transfer}) which fully encode the
relation between underlying true sky power and observed power for both
pipelines.

In both analyses, we estimate the covariance matrix of our
power spectrum estimates from the scatter found in the power spectra
measured from simulations containing both signal and
noise:
\be
\lgl \Delta {\bf P}_b \Delta {\bf P}_{b'}\rgl = 
\lgl({\bf P}_b - \overline{\bf P}_b)({\bf P}_{b'} - \overline{\bf P}_{b'})
\rgl_{MC} \, ,
\label{eqn:band_powers_covar}
\ee
where $\overline{\bf P}_b$ denotes the average of each band power over
all simulations. Note finally that the covariance properties of the
power spectra estimates are dependent on whether the correction for
mode-mixing induced by the sky cut is applied or not. We return to
this issue in Section~\ref{sec:final_spectra} where we compare the
results from our two analyses.

\subsection{Simulations}
\label{sec:sims}
In simulating QUaD, we follow the procedure described in
Section 5 of Paper II with some important differences, which we now
discuss. Algorithmically both of our analysis pipelines adopt the same
approach to creating simulated timestreams and only differ in the
final map-making stage as described in Section
\ref{sec:template_removal}.

\subsubsection{Signal simulations}
\label{sec:sig_sims}
To create the signal component in the simulations, we first generate
model $TT, EE, TE$ and $BB$ CMB power spectra using CAMB \citep{lewis00}. 
The input cosmology consists of the best-fitting
$\Lambda$CDM model to the 5-year WMAP data set \citep{dunkley09}. Note that the model
spectra used include the effects of CMB lensing and so the
input $B$-mode power is non-zero. (For comparison, in Paper II, our input
model was the best-fit to the 3-year WMAP data set and our input
$B$-mode power was set to zero.)

Realizations of CMB skies are then generated from these model spectra
using a modified version of the HEALPix software. These maps are
generated at a resolution of 0.4 arcmin ($N_{\rm side} = 8192$). The
simulated maps are then projected onto a 2D cartesian grid and
convolved with the beam model for each detector channel. The
resolution used for this intermediate map is 0.6 arcmin. The
generation of the sky maps for each detector and deck angle and
interpolation to simulated TOD then proceeds exactly as described in
Section 5.1 of Paper II.

\subsubsection{Noise simulations}
\label{sec:noise_sims}
In order to simulate realistic noise, we must first measure the noise
properties from the real data. However, the undifferenced data
contains not only noise but also CMB signal and ground signal 
(see equation \ref{eqn:tod_model1}). The instantaneous signal-to-noise
in the timestream is negligible and so the CMB component can be
safely ignored. However, the same is not true for the ground signal which, in
some cases, is a very significant component in the timestream,
particularly in polarization. The data must therefore be cleaned
of the ground component before measuring the noise power spectra. 

One might think that the best way to achieve this would be to measure
the noise spectra from data which has been template subtracted using
equation (\ref{eqn:template_subtract}). Such a procedure could, in
principle, be iterated and is similar to procedures suggested for
measuring the noise properties of CMB data when the signal component
is non-negligible (e.g. \!\citealt{ferreira00}). However, as noted in
Section \ref{sec:template_removal}, our template removal procedure
filters the noise in a non-trivial fashion. In particular,
because of the non-uniform azimuth coverage of the scan strategy, the
ground templates are noisier at each end than they are in the central
regions. The result is that after subtracting the templates, the noise
is no longer uniform and this prohibits its characterization through
simple FFT-based power spectrum estimators --- since the noise is no
longer a stationary Gaussian random process, a power spectrum
description will fail.
 
To avoid these complications, we have measured the noise properties of
the TOD from data which has been lead-trail differenced. The
differencing efficiently removes the ground signal while under the
assumption that the noise is stationary over a $30$ minute timescale,
the power spectra of the \emph{undifferenced} TOD are simply the
spectra measured from the differenced TOD divided by 2. For each pair
of lead-trail observations, we therefore assign the power spectra
measured from the differenced data, appropriately normalized, to each
of the lead and trail scan-sets. Simulated noise-only timestreams are
then generated exactly as described in Section 5.2 of Paper II.

Examining the QUaD data and comparing it to simulated data obtained
using the above process, there are occasions where our assumption of
stationarity over a $30$ minute timescale is not satisfied. However,
for the majority of the data the assumption is good and it is only
ever a poor one for our temperature analysis. Moreover, a thorough
comparison of the statistics of the simulated and real data indicates
that our procedure provides an excellent description of the noise
properties of the undifferenced data when averaged over each day for both
temperature and polarization. Further averaging over tens of pixels
and hundreds of observation dates will result in these rare failures of
our noise model having a negligible impact on the results. 

Once generated, both the signal-only and noise-only simulated TOD are
processed into $T, Q$ and $U$ maps in an identical manner to that used
for the real data. In particular, note that both the ground template
subtraction and polynomial filtering are applied also to the simulated
data and so the effects of filtering on both the signal and on the
noise are fully accounted. Finally, to obtain simulated maps
containing both signal and noise, we simply add the signal-only and
noise-only maps. Since all of our data processing steps are linear
operations, this final step results in simulated maps no different to
those which would have been obtained if we had instead summed the
signal-only and noise-only TODs, and is computationally more
efficient.

\subsection{Transfer functions and band power window functions}
\label{sec:transfer}
We estimate the transfer function from our suite of signal-only
simulations. In the absence of noise, the mean of the recovered
pseudo-$C_\ell$ spectra will equal their expectation values,
\be
\lgl \widetilde{\bf C}_\ell \rgl_{MC} = \sum_\ld {\bf M}_\lld F_\ld
     {\bf C}_\ld \, ,  
\label{eqn:expvals_sig}
\ee 
where ${\bf C}_\ell$ are the input model spectra used to create the
simulations. For a small-area survey such as QUaD, the unbinned
coupling matrix, ${\bf M}_\lld$, is singular and so
equation~(\ref{eqn:expvals_sig}) cannot be solved directly. In Pipeline 1, 
we iteratively solve this equation to provide an estimate of
$F_\ell$. With a reasonable starting guess, convergence is typically
reached in just a few iterations. For Pipeline 1, the band power window
functions, ${\bf W}_{b \ell}$, defined by \citep{knox99}, 
\be
\lgl {\bf P}_b \rgl = \sum_\ell \frac{{\bf W}_{b \ell}}{\ell}
  \frac{\ell(\ell + 1)}{2 \pi} \lgl {\bf C}_\ell \rgl   \, ,
\ee
are given by 
\be
\frac{{\bf W}_{b \ell}}{\ell} = \frac{2 \pi}{\ell(\ell + 1)} F_\ell \, \sum_{b'} {\bf K}^{-1}_{bb'} \sum_\ld P_{b' \ld}
{\bf M}_\ldl \, .
\ee

Pipeline 2 calculates its band power window functions numerically as
described in Section 6.6 of Paper II. In order to calculate the
transfer functions, Pipeline 2 simply takes the ratio of the
mean band powers recovered from signal-only simulations and the
expectation values for each band power:
\be
F_b = \frac{\sum_\ell P_{b \ell} \lgl \widetilde{\bf C}_\ell
  \rgl_{MC}}{ \lgl {\bf P}_b \rgl } \, .
\ee

\begin{figure}[t]
  \vspace{0.3cm}
  \centering
  \resizebox{0.48\textwidth}{!}{  
    \rotatebox{-90}{\includegraphics{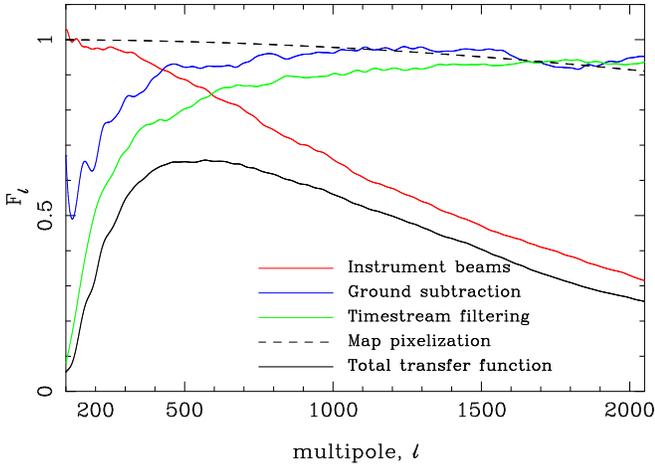}}}
  \caption{Total transfer function, $F_\ell$ (black curve) as measured
  in our Pipeline 1 analysis for the 150~GHz channel. Also shown are
  the transfer functions for timestream filtering (green),
  ground-template removal (blue) and beam suppression (red) in
  isolation. The dashed curve shows the suppression of signal due to
  map pixelization for a HEALPix resolution, $N_{\rm side} =
  2048$. The total transfer function is the product of these four
  individual curves.}
  \label{fig:transfer}
\end{figure}

Figure~\ref{fig:transfer} shows the derived transfer function from
Pipeline 1. (The transfer function from Pipeline 2 is similar.)  As
mentioned earlier, this function encapsulates all effects due to
timestream filtering, beam suppression and the filtering due to
removal of the ground templates. To demonstrate the relative size of
these effects, we also plot the transfer function derived from special
simulations with these three effects included in isolation. Of
particular interest is the transfer function describing the
ground-removal procedure --- this curve, in effect encapsulates the
lossiness of the technique. On scales where this curve
is $\gsim 0.5$, we gain an improvement over the field-differencing
technique. We see that below $\ell \sim 200$, there is little gain
from our template-removal technique over field-differencing. The
amount of signal retained then climbs rapidly and is effectively unity
by $\ell = 1000$.

\begin{figure}[t]
  \vspace{0.3cm}
  \centering
  \resizebox{0.48\textwidth}{!}{  
    \rotatebox{-90}{\includegraphics{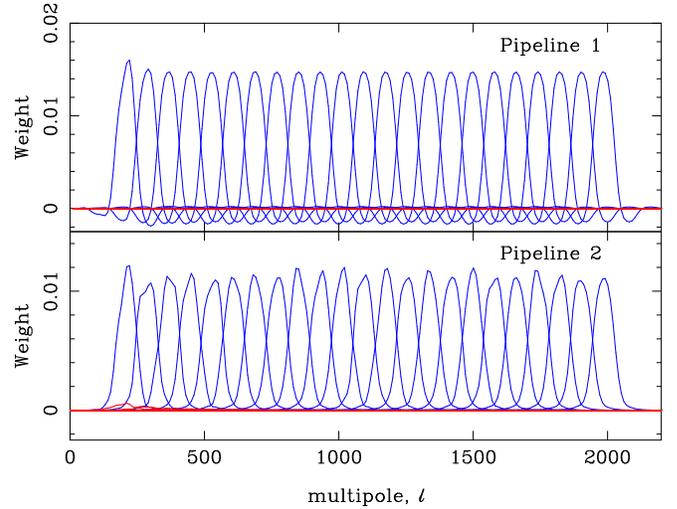}}}
  \caption{Band power window functions for Pipeline 1 (top panels) and
  Pipeline 2 (bottom panels) for the $C^{BB}_\ell$ power spectrum. The blue
  curves show the response to true $BB$ power and the red lines show
  the response to $EE$ power.}
  \label{fig:bpwfs}
\end{figure}

Figure~\ref{fig:bpwfs} shows the band power window functions (BPWFs),
plotted as $W_{b \ell} / \ell$, for $C^{BB}_\ell$ from both our
pipelines for the 150~GHz channel. These functions describe the
response of our band power measurements to the true sky signal at each
multipole.  The negative wings in the BPWFs of Pipeline 1 are a direct
result of the application of the correction for mode-mixing effects as
described in Section~\ref{sec:spectra_estimation}.

\begin{figure}[t]
  \vspace{0.3cm}
  \centering
  \resizebox{0.48\textwidth}{!}{  
    \rotatebox{-90}{\includegraphics{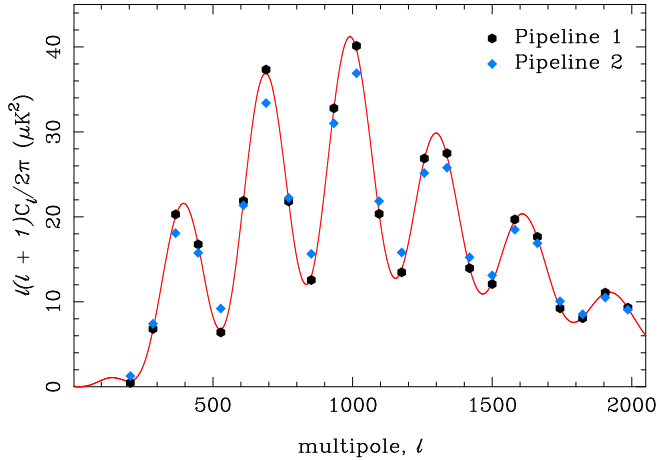}}}
  \caption{Expectation values for the $\Lambda$CDM $E$-mode power
  spectrum for both pipelines at 150~GHz. The mapping of true sky
  power to observed band powers for both pipelines is fully encoded in
  their associated band power window functions
  (Figure~\ref{fig:bpwfs}).}
  \label{fig:expvals}
\end{figure}

The expectation values for our $EE$ band powers assuming the best-fit
$\Lambda$CDM model to the WMAP 5-year data are shown in
Figure~\ref{fig:expvals}. Note that the apparent improvement in going
from Pipeline 2 to Pipeline 1 in terms of the agreement between the
true sky power and the band power expectation values does not come
without a price --- as a result of applying the mode-mixing
correction, the error bars for Pipeline 1 are enhanced with respect to
the Pipeline 2 errors. The covariance properties also change between
the two analyses such that the total information content is preserved.

We emphasize that, in terms of either the accuracy or the precision of
the recovery of true sky power, neither analysis is superior. This is
clear from the fact that one can transform between the band power
estimates and covariances of the two analyses via a simple and exact
matrix operation. Whether to apply the mode-mixing correction is
simply a matter of preference and is only relevant for visual
interpretation of the results --- one can choose to have smaller error
bars or one can choose to have band powers which better trace the
underlying true sky power but one cannot have both.

\subsection{Power spectrum results}
\label{sec:spectra_results}
\begin{figure*}[t]
  \vspace{0.3cm}
  \centering
  \resizebox{0.9\textwidth}{!}{  
    \rotatebox{0}{\includegraphics{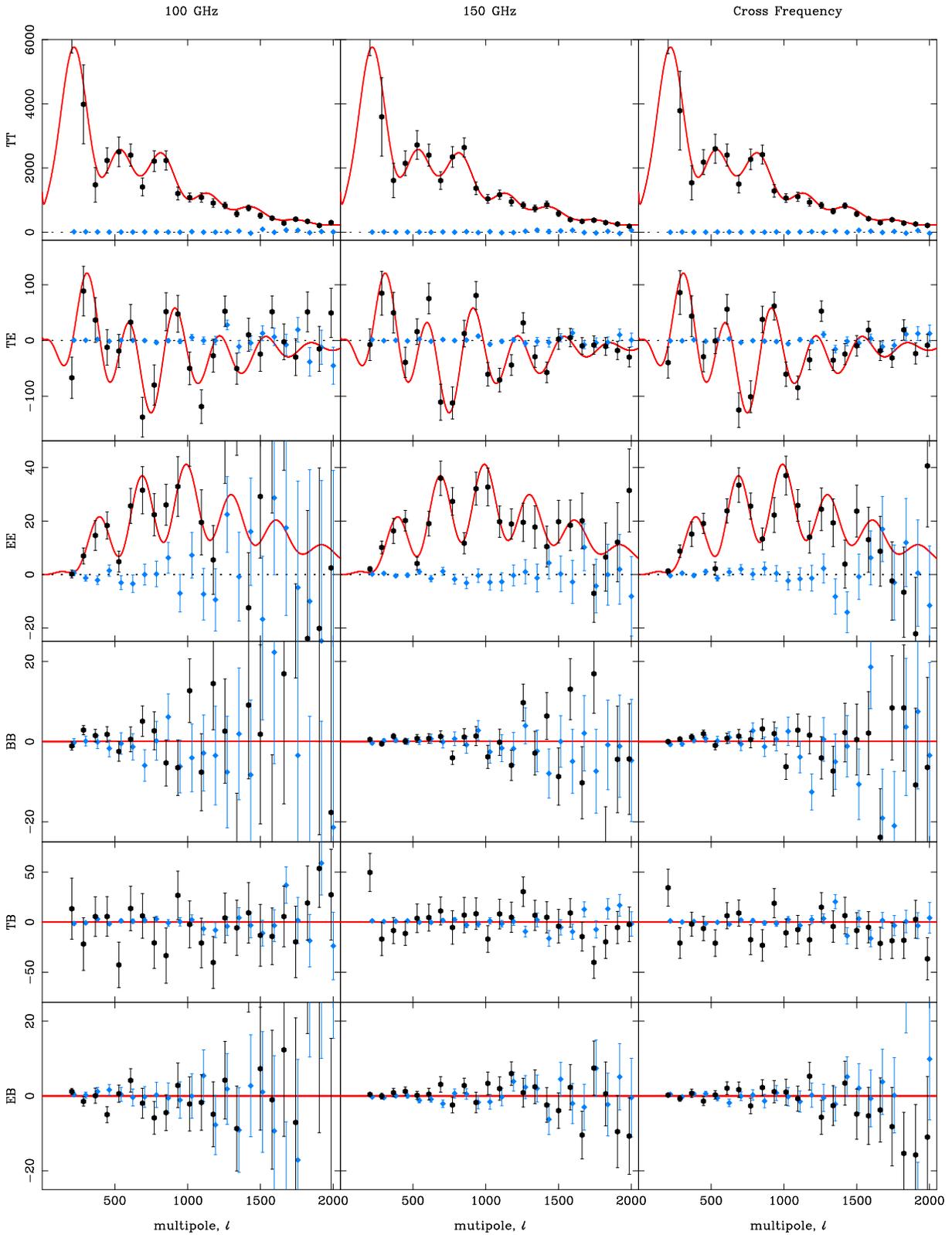}}}
  \caption{Full set of power spectrum results from Pipeline 1 where the
  quantity plotted in the y-direction is $\ell (\ell + 1) C_\ell / 2 \pi$. The red
  curves are the spectra predicted by the best-fit $\Lambda$CDM model
  to the WMAP 5-year data \citep{dunkley09}. The blue points show the
  power spectra measured from our deck angle differenced maps and
  represent a stringent test for residual ground contamination --- see
  the text of Section~\ref{sec:jackknives} for details.}
  \label{fig:all_spectra}
\end{figure*}

We apply the procedures described in
Section~\ref{sec:spectra_estimation} to estimate the six possible CMB
power spectra ($TT, TE, EE, BB, TB, EB$) from the QUaD maps. In
addition to the 100 and 150~GHz auto-spectra, we also estimate the
100-150~GHz cross-spectra as described in Paper II. In the case where
the noise is uncorrelated between the two frequency channels, the
cross spectra do not require the noise-debiasing step. Although, in
practice, we do apply this correction, the correction is modest for
the cross spectra so these measurements will be much less sensitive to
the details of our noise model. The power spectrum results from
Pipeline 1 are presented in Figure~\ref{fig:all_spectra}. We make strong
detections of the $TT$, $TE$ and $EE$ spectra which show good
agreement with those predicted by the best-fitting
$\Lambda$CDM model to the 5-year WMAP results. The $BB$, $TB$ and $EB$
spectra are consistent with null within the noise. 
The results from Pipeline 2 show a similar agreement with the $\Lambda$CDM
model. 

\subsection{Jackknife tests}
\label{sec:jackknives}
We have subjected our analysis to the same set of systematic
tests as performed in Paper II. These tests involve splitting the data
into two roughly equal parts. Maps made from the two data subsets are
subtracted and the power spectra of the residual maps are
calculated. Any deviation of these ``jackknife'' spectra from 
null would indicate systematic contamination in the
data. For a detailed description of the various ways in which we split
the data, we refer the reader to Section 7 of Paper II. 

The strongest test is the so-called ``deck jackknife'' test
where we split the data according to the boresight rotation angle
(deck angle) of the observations. This test, in particular, will be
strongly sensitive to any residual ground contamination remaining in
the data after applying the procedure described in
Section~\ref{sec:map_making}. In Figure~\ref{fig:all_spectra}, the
deck jackknife spectra are plotted alongside the spectra measured from
the undifferenced maps. It is clear from this figure that the power
measured in the deck-differenced maps is small compared to the
measured signals in $TT$, $TE$ and/or $EE$. The other data splits
which we consider are splitting the data according to orientation of
the PSB pairs on the focal plane (see Section~\ref{sec:obs_lowlevel}),
a split between the forward and backward scans and a split in
observation dates which roughly separates the data into the 2006 and
2007 observing seasons. We also take the difference of the 100 and
150~GHz maps. This ``frequency jackknife'' test is not strictly a test
for systematic effects in the data. Rather, it is a strong test for
``foreground'' (i.e. \!non-CMB) astrophysical emission. 

The cancellation of the signal apparent in
Figure~\ref{fig:all_spectra} is visually impressive. To investigate
whether the differenced spectra are formally consistent with null, as
in Paper II, we have performed $\chi^2$ tests against the null
model. Note that we compare this statistic with the $\chi^2$
distribution as measured from our simulations rather than against the
theoretical $\chi^2$ curve, calculating the ``probability to exceed''
(PTE) the observed value by random chance. Low numbers therefore
indicate a problem. The PTE values for each of our measured spectra
(from Pipeline 1) are presented in Table~\ref{tab:pte_stats}.

\begin{deluxetable}{c c c c c c c}
\tablecaption{Jackknife PTE values from $\chi^2$ tests}
\tablehead{\colhead{Jackknife} &
\colhead{$TT$} & \colhead{$TE$} & \colhead{$EE$} &
\colhead{$BB$} & \colhead{$TB$} & \colhead{$EB$} }
\startdata
Deck angle:     &        &        &        &        &        &       \\
100 GHz         & 0.000  & 0.415  & 0.883  & 0.933  & 0.598  & 0.917 \\
150 GHz         & 0.008  & 0.295  & 0.963  & 0.988  & 0.258  & 0.423 \\
Cross           & 0.000  & 0.028  & 0.780  & 0.197  & 0.287  & 0.527 \\
                &        &        &        &        &        &       \\
Scan direction: &        &        &        &        &        &       \\
100 GHz         & 0.008  & 0.017  & 0.122  & 0.812  & 0.478  & 0.518 \\
150 GHz         & 0.080  & 0.665  & 0.755  & 0.153  & 0.515  & 0.485 \\
Cross           & 0.000  & 0.608  & 0.155  & 0.783  & 0.487  & 0.263 \\
                &        &        &        &        &        &       \\
Split season:   &        &        &        &        &        &       \\
100 GHz         & 0.743  & 0.287  & 0.350  & 0.655  & 0.840  & 0.413 \\
150 GHz         & 0.000  & 0.387  & 0.242  & 0.022  & 0.340  & 0.647 \\
Cross           & 0.273  & 0.065  & 0.110  & 0.160  & 0.630  & 0.850 \\
                &        &        &        &        &        &       \\
Focal plane:    &        &        &        &        &        &       \\
100 GHz         & 0.173  & 0.872  & 0.690  & 0.813  & 0.703  & 0.672 \\
150 GHz         & 0.530  & 0.397  & 0.910  & 0.988  & 0.933  & 0.715 \\
Cross           & 0.270  & 0.012  & 0.493  & 0.105  & 0.735  & 0.578 \\
                &        &        &        &        &        &       \\
Frequency       & 0.000  & 0.362  & 0.418  & 0.588  & 0.208  & 0.783 \\
difference      &        &        &        &        &        &       \\
\enddata
\label{tab:pte_stats}
\end{deluxetable}

Examining the table, the PTE values for all the spectra bar $TT$
reveal no significant problems, the numbers being consistent with a
uniform distribution between zero and one. In contrast, many of our
$TT$ jackknife spectra are clearly inconsistent with a null
signal. The failure is perhaps excusable in the case of the frequency
difference (since there are astrophysical reasons why this test might
fail) but taken as a whole, the statistics for $TT$ (and to a lesser
extent, some of the $TE$ numbers) suggest that there is some degree of
residual systematics present in our temperature maps. The PTE
statistics for Pipeline 2, although not identical, show the same
general pattern of jackknife failures for the $TT$ spectra. Comparing
to our previous results, there were hints of a problem with the $TT$
jackknife tests in Paper II but to a lesser extent than is apparent
now. This is possibly due to the fact that with our increased
sensitivity we can measure both the signal and systematics to greater
precision. Alternatively, it could indicate that the template removal
procedure is not as effective as field-differencing in removing the
ground. 

\begin{figure}[h]
\begin{center}
\resizebox{0.48\textwidth}{!}{
  \rotatebox{-90}{\includegraphics{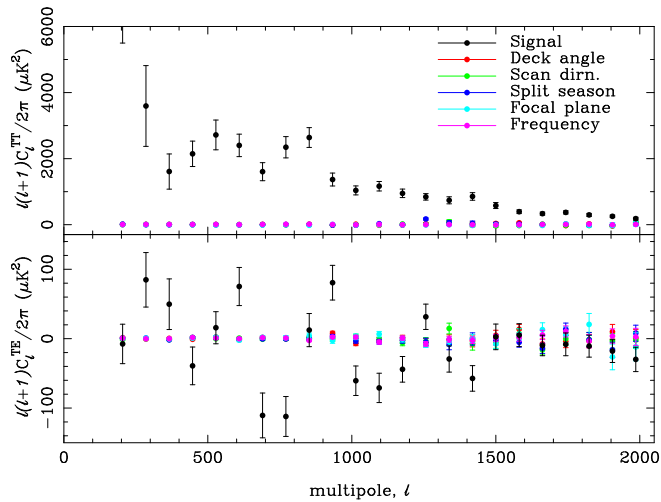}}}
\end{center}
\caption{Signal and jackknife $TT$ (top panel) and $TE$ (bottom panel)
  power spectra for the 150~GHz channel. The black points show the
  measured signal spectra and the colored points are the jackknife
  spectra. The levels of power measured in the jackknife maps are
  negligible when compared to the sample variance driven errors on the
  signal spectra.} 
\label{fig:tt_te_jacks}
\end{figure}

These jackknife failures indicate that residual systematic effects in
our temperature maps are significant with respect to the errors on the
jackknife spectra. However, the jackknife errors contain only noise
whereas the errors on our measured $TT$ and $TE$ CMB power spectra
also contain considerable sample variance. To assess how significant
the residual systematics are, in Figure~\ref{fig:tt_te_jacks}, we plot
the measured $TT$ and $TE$ jackknife band powers alongside the signal
band powers for the 150~GHz channel. Clearly, when compared to the
sample variance dominated signal spectra, the degree of residuals are
much less significant. Repeating the $\chi^2$ analysis using the
signal covariance matrices in place of the jackknife covariance
matrices, we find that the residual contamination is negligible. In
summary, although our TT jackknife tests indicate the presence of
residual systematics, they also clearly demonstrate that these
residuals are irrelevant compared to both the measured sky signal and
its associated sample variance. Note this is also true for the levels
of foreground contamination in our maps since our frequency difference
test is sensitive to foregrounds.

\subsection{Examination of final power spectra}
\label{sec:final_spectra}
\begin{figure*}[t]
  \vspace{0.3cm}
  \centering
  \resizebox{0.9\textwidth}{!}{  
    \rotatebox{-90}{\includegraphics{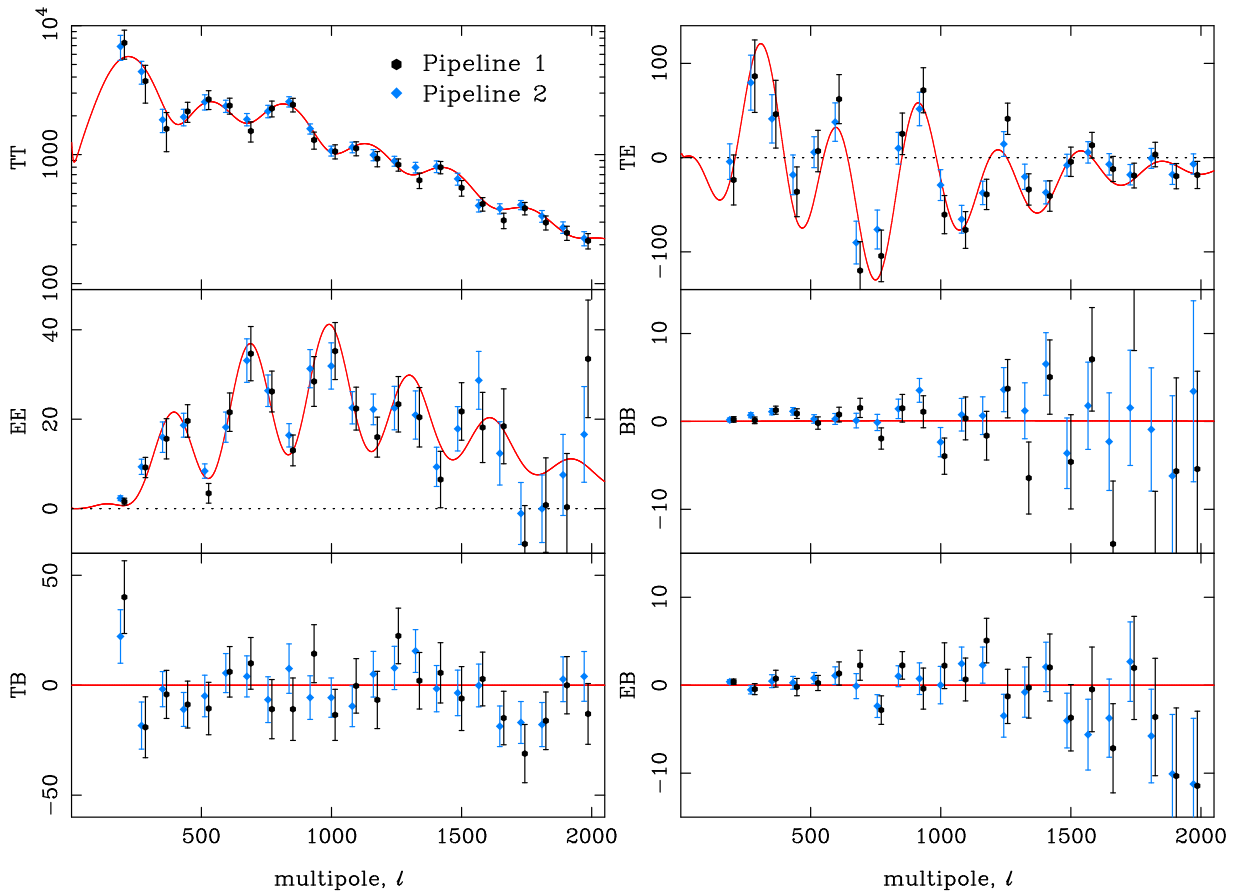}}}
  \caption{Final power spectrum results obtained from an optimal
  combination of the 100~GHz, 150~GHz and cross-frequency power
  spectra. Spectra from both pipelines (again, plotted as 
  $\ell (\ell + 1) C_\ell / 2 \pi$) are shown in comparison to the expected
  spectra in the concordance $\Lambda$CDM model. Note the $TT$ power
  spectrum is plotted with a log-scale in the y-axis. For clarity, the
  two sets of points have been slightly offset in the horizontal
  direction.}
  \label{fig:combined_spectra}
\end{figure*}

To produce a set of final power spectra, we combine the
single-frequency and 100-150~GHz cross spectra shown in
Figure~\ref{fig:all_spectra} following the procedure described in
Section 9 of Paper II. The final estimates of the power spectra are
shown in Figure~\ref{fig:combined_spectra} for both of our analysis
pipelines. Once again, the model curves plotted are the best-fitting
$\Lambda$CDM model to the WMAP 5-year results. We have performed
$\chi^2$ tests of all spectra and both analyses show perfectly
acceptable agreement with this model and with each other. Pipeline 1
$\chi^2$ and PTE values are given in Table~\ref{tab:signal_stats}.

\begin{deluxetable}{c c c}
\tablecaption{$\chi^2$ tests of the combined spectra against the
  $\Lambda$CDM and null models.}
\tablehead{\colhead{ } & \colhead{vs. $\Lambda$CDM} & \colhead{vs. null}}
\startdata

$\chi^2$ for 23 d.o.f.: &       &        \\
$TT$       & 15.71 & 1512.86             \\
$TE$       & 27.10 &  116.38             \\
$EE$       & 24.87 &  413.81             \\
$BB$       & 37.00 &   37.87             \\
$TB$       & 28.99 &    ---              \\
$EB$       & 22.25 &    ---              \\
                                         \\
PTE values: &       &                    \\
$TT$       & 0.830 & < machine precision \\
$TE$       & 0.208 & $7.9\times10^{-15}$ \\
$EE$       & 0.303 & < machine precision \\
$BB$       & 0.024 &            $0.019$  \\
$TB$       & 0.145 &               ---   \\
$EB$       & 0.445 &               ---   \\

\enddata
\label{tab:signal_stats}
\end{deluxetable}

\begin{figure*}[t]
  \vspace{0.3cm}
  \centering
  \resizebox{0.9\textwidth}{!}{  
    \rotatebox{-90}{\includegraphics{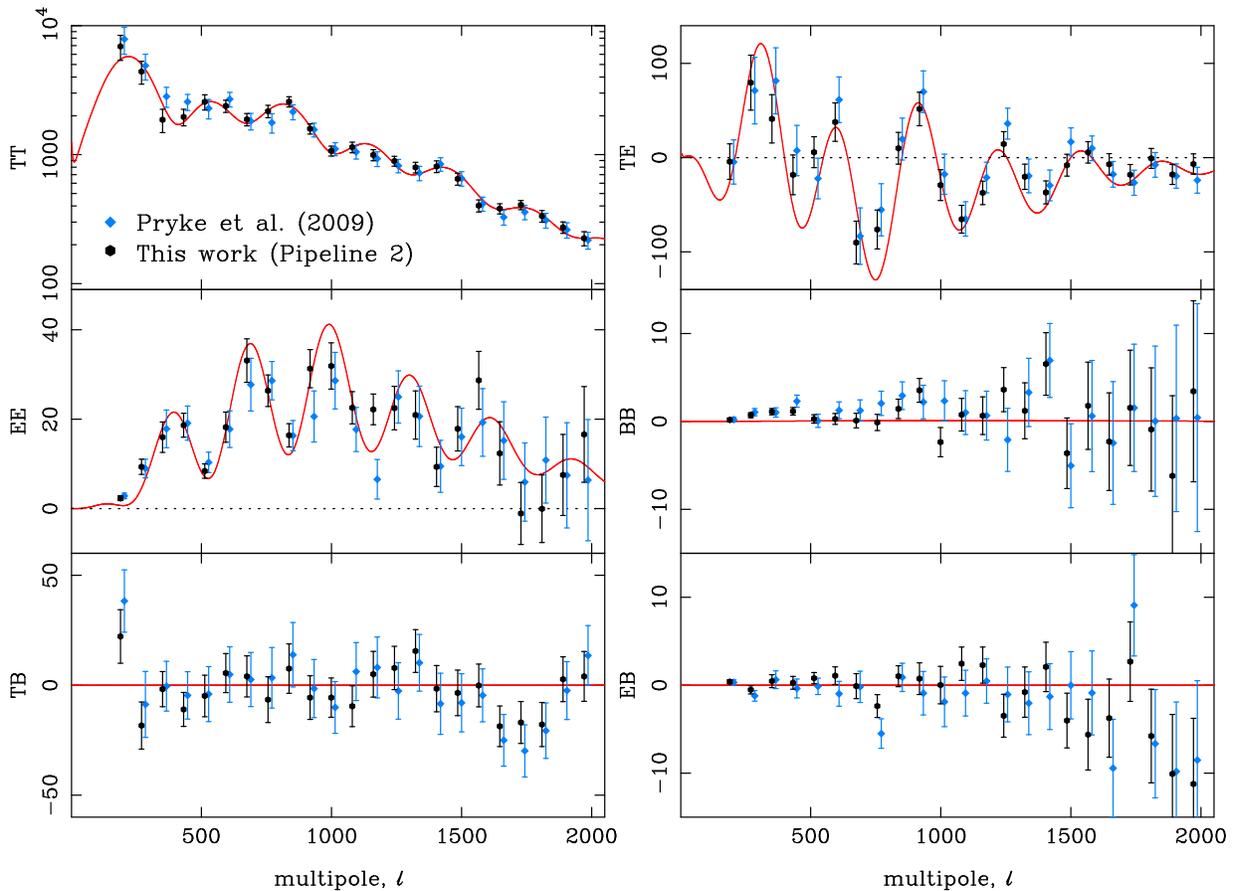}}}
  \caption{Comparison of our final power spectra from Pipeline 2 with the
  field-differenced analysis of \cite{pryke09}. The uncertainties
  on our band power measurements have been reduced by $\sim 30\%$ and
  our new beam models result in a small increase in amplitude in the measured
  spectra for multipoles, $\ell \gsim 700$. For clarity, the two sets of
  points have been slightly offset in the  horizontal direction.}
  \label{fig:fd_nfd_compare}
\end{figure*}

Comparing the two sets of results presented in
Figure~\ref{fig:combined_spectra} we see that the nominal error-bars
are smaller for Pipeline 2 and that the Pipeline 2 points appear to
trace a slightly smoothed version of the Pipeline 1 points. Both of
these effects are simply a result of the differing band power
window functions as discussed in Section~\ref{sec:transfer}. Note that
neighboring band powers in Pipeline 1 are $\sim 10 \%$ anti-correlated
whereas they are $\sim 10 \%$ positively correlated in Pipeline
2. Correlations between non-adjacent band powers are negligible for
both analyses.

Figure~\ref{fig:fd_nfd_compare} shows a comparison with our previous
results from Paper II. Note that we perform this comparison using
Pipeline 2 since this pipeline is an extension of the analysis
presented in Paper II and so is directly comparable. Two effects are
apparent in this figure. First, the uncertainties on all of our power
spectra have been reduced by $\sim 30\%$ as a result of the increase
in sky area afforded by our template-based ground removal
technique. Second, implementing the improved beam models described in
Section~\ref{sec:beam_model} has resulted in a slight increase in the
amplitude of our power spectrum measurements for multipoles, $\ell
\gsim 700$. This impacts mostly on the high signal-to-noise
measurements of the $TT$ spectrum on small scales.

\begin{figure*}[t]
  \vspace{0.3cm}
  \centering
  \resizebox{0.9\textwidth}{!}{  
    \rotatebox{-90}{\includegraphics{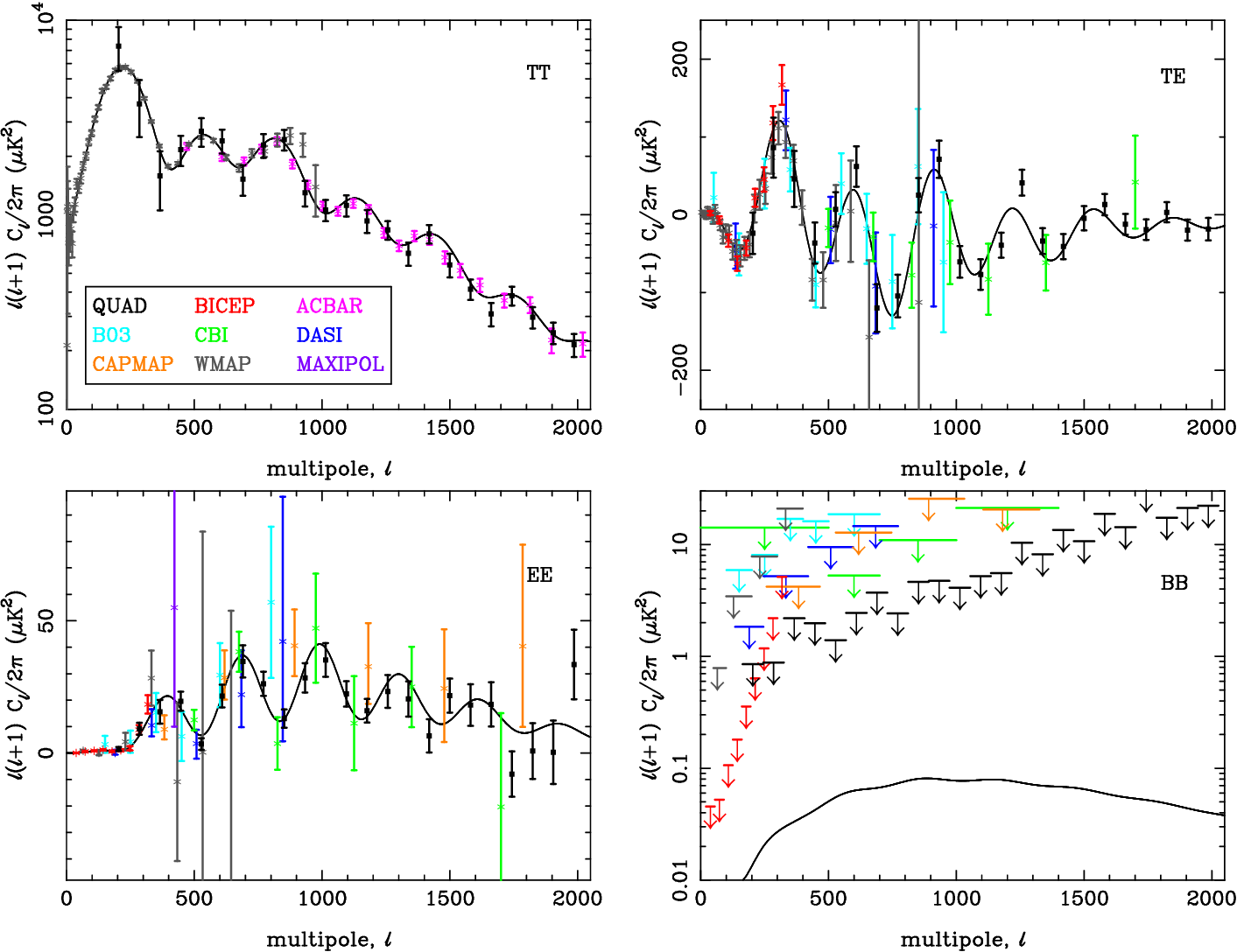}}}
  \caption{QUaD measurements of the $TT$, $TE$ $EE$ and $BB$ power spectra
    compared to results from the WMAP \citep{nolta09}, ACBAR
    \citep{reichardt09}, BICEP \citep{chiang09}, B03 \citep{piacentini06, montroy06}, CBI
    \citep{sievers07}, CAPMAP \citep{bischoff08}, MAXIPOL \citep{wu07} and DASI
    \citep{leitch05} experiments. The $BB$ measurements are plotted as
    95\% upper limits. The smooth black curves in each panel are the
    power spectra expected in the best-fit $\Lambda$CDM model to the
    WMAP 5-year data.} 
  \label{fig:compare_spectra}
\end{figure*}

Figure~\ref{fig:compare_spectra} shows our measured power spectra 
from Pipeline 1 in comparison with the published results from a number
of other CMB experiments. 

The QUaD power spectra data, along with the associated band power
covariance matrices and band power window functions (for both of our
analysis pipelines) are available for download at
http://quad.uchicago.edu/quad.

\subsection{Acoustic oscillations in the $E$-mode power spectrum}
\begin{figure}[t]
  \vspace{0.3cm}
  \centering
  \resizebox{0.45\textwidth}{!}{  
    \rotatebox{-90}{\includegraphics{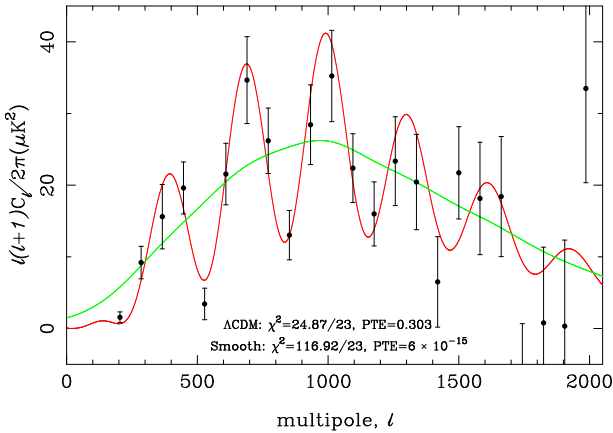}}}
  \caption{QUaD's measurements of the $EE$ spectrum (black points)
  compared to the $\Lambda$CDM model (red curve) and a model without
  peaks (green curve). The data are incompatible with the no-peak
  scenario --- the probability that the smooth curve is correct is $<
  10^{-14}$.}
  \label{fig:ee_smooth_lcdm_test}
\end{figure}

We have assessed the significance with which we detect the acoustic
oscillations in the $EE$ power spectrum by repeating the analysis of
Section 9.3 in Paper II. For this test, we compare our $EE$
measurements against both the $\Lambda$CDM model and against a
heavily smoothed version of the $\Lambda$CDM curve. The results of
this test are shown in Figure~\ref{fig:ee_smooth_lcdm_test}. The QUaD
detection of acoustic oscillations in the $E$-mode power spectrum is
now beyond question -- the probability that the true $E$-mode spectrum is a
smooth curve has dropped from $0.001$ with our previous analysis to $<
10^{-14}$. We have also repeated our Paper II analysis
where we used ``toy models'' of the $E$-mode spectrum to fit the peak
spacing, phase and amplitude of the acoustic oscillations. With our
previous measurements, we constrained the peak spacing to be $\Delta
\ell_s = 306 \pm 10$, the phase to be $\phi = 13^\circ \pm 33^\circ$
and the amplitude to be $a = 0.86 \pm 0.17$. Repeating the analysis
with our new measurements, we find $\Delta \ell_s = 308 \pm 7$, $\phi
= 6^\circ \pm 22^\circ$ and $a = 0.96 \pm 0.10$. For comparison,
when we perform the analysis with the QUaD band power values replaced by
their expectation values in the $\Lambda$CDM model, we
find $\Delta \ell_s = 310$, $\phi = 13^\circ$ and $a = 0.99$. 
These results confirm that the polarization peak spacing and the phase
relationship between the temperature and polarization acoustic
oscillations are as expected in the $\Lambda$CDM model. Passing this
(non-trivial) test further strengthens the foundations of this model. 

\section{Cosmological interpretation}
\label{sec:parameters}
The QUaD results presented in this paper are the
most precise determination of the CMB polarization, and of its
correlation with the CMB temperature, at angular scales $\ell > 200$,
to date. Within the standard $\Lambda$CDM model, given a precise
measurement of the $TT$ power spectrum (e.g. from WMAP), the
$TE$ and $EE$ spectra on all but the very largest scales are
deterministically predicted. Nevertheless, sufficiently accurate
measurements of these spectra can still tighten constraints
on cosmological parameters. In particular, precise measurements of the
$TE$ and $EE$ acoustic peaks and valleys can help constrain the
cosmological parameters which determine this peak pattern. 

Beyond the standard $\Lambda$CDM model, power spectrum measurements
from small-scale experiments such as QUaD can add further
information. Two such extensions to the $\Lambda$CDM model which are
well-motivated in the context of single-field slow-roll inflation
models are a possible tensor component in the primordial perturbation
fields and a scale dependence (or ``running'') in the scalar spectral index. Placing
constraints on tensor modes through measurements of the large-scale
$B$-mode polarization induced by a gravitational wave background from
inflation is a major goal of ongoing and future CMB polarization
experiments. Although the scales probed by QUaD's polarization
measurements are too small to constrain this $B$-mode signal directly, 
QUaD can help through its ability to constrain the scalar spectra
index, $n_s$, since this parameter is correlated with the
tensor-to-scalar ratio. For a $\Lambda$CDM model extended to include a
running in $n_s$, measurements of the high-$\ell$ temperature power
spectrum can be useful in constraining both $n_s$ and the degree
of running.  

In this section, we constrain cosmological models by adding the QUaD
temperature and polarization data (i.e. \!our new measurements of the
$TT$, $EE$, $TE$ and $BB$ spectra) to the results of two other CMB
experiments --- the WMAP 5-year analysis \citep{nolta09} and the final
results from the ACBAR experiment \citep{reichardt09}. We will also
investigate the effect of adding large-scale structure data by
including measurements of the present-day matter power spectrum,
$P(k)$ from the SDSS Luminous Red Galaxy (LRG) 4th data release
\citep{tegmark06}.

In this paper, we focus on what the QUaD dataset taken as a whole adds
to parameter constraints. In addition to investigating further
extensions to the $\Lambda$CDM model, we will fully explore the
consistency of our temperature-only and polarization-only parameter
constraints in a future paper (Gupta et al., in prep). See also
\cite{castro09} for temperature-only and polarization-only constraints
obtained using our previous power spectrum results of Paper II.

\subsection{Methodology}
\label{sec:parameters_methods}
To obtain our constraints, we perform a Monte-Carlo Markov Chain
(MCMC) sampling of the cosmological parameter space. To do this, we
use the publicly available CosmoMC package \citep{lewis02}, which in
turn uses the CAMB code \citep{lewis00} to generate the CMB and matter
power spectra.\footnote{%
Note that we have used the June 2008 version
of the CAMB software. This version included a revised model of the
reionization history as compared to previous versions of CAMB. In
particular, the mapping between the optical depth, $\tau$ and the
reionization redshift, $z_{\rm re}$ changed at the $\sim 10\%$ level
--- see~\cite{lewis08} for details. This should be borne in mind when
comparing our results to previous analyses such as \cite{dunkley09}
and \cite{reichardt09} who used pre-March 2008 versions of the CAMB
software.}

We make use of the publicly available WMAP likelihood software from the
LAMBDA\footnote{http://lambda.gsfc.nasa.gov/} website. We marginalize
over a possible contribution to the temperature power spectrum from
the Sunyaev-Zel'dovich (SZ) effect following the WMAP analysis
\citep{dunkley09}. To do this, we use the SZ templates from \cite{komatsu02} (also
available from the LAMBDA website) and the known frequency dependence
of the SZ effect. In order to avoid possible contamination
from residual point sources, we exclude the ACBAR band powers above
$\ell = 2000$. For the same reason, we do not include our own $TT$
measurements at $\ell > 2000$ \citep{friedman09}.

Marginalization over the WMAP beam uncertainty is included in the WMAP
likelihood code and we also marginalize over the quoted ACBAR
calibration and beam uncertainties. We take the latter to be
a 2\% error on a 5 arcmin (FWHM) Gaussian beam as assumed in the ACBAR
CosmoMC data file. For QUaD, we marginalize over our 3.5\% calibration
uncertainty and over the uncertainty in our beam. As described in
Appendix~\ref{app:beam_model_app}, our beam uncertainties are
dominated by uncertainties in the level of our sidelobes rather than
in the effective FWHM of our main lobe beams. We therefore 
marginalize over the full $\ell$-dependent beam uncertainty shown in
Figure~\ref{fig:beam_uncert}. Where we include the SDSS LRG
data, we marginalize over both the amplitude of the matter power
spectrum and over a correction for scale-dependent non-linear density
evolution using the methods described in \cite{tegmark06}. 

We model the likelihood functions for the QUaD auto-spectra as offset
log-normal distributions \citep{bond00}. The required noise-offsets
are derived from our signal-only and noise-only simulations. (We model
the $TE$ likelihood as a Gaussian distribution.) We include all
covariances apparent (above the numerical noise) in our
simulation-derived covariance matrix (equation
\ref{eqn:band_powers_covar}). In addition to same-spectrum
covariances, this includes non-zero $TT$-$TE$, $EE$-$TE$,
$TT$-$EE$ and $EE$-$BB$ correlations.

Note that, for our main MCMC analysis, we do not include our
measurements of the parity-violating spectra, $TB$ and $EB$ since
these spectra are expected to vanish in standard $\Lambda$CDM models
and its usual variants. However, in Section~\ref{sec:params_parviol},
we will use these spectra to constrain possible parity-violating
interactions to the surface of last scattering (see
e.g.~\citealt{lue99}) following our previous work \citep{wu09}.

Finally, in Section~\ref{sec:params_bmode_limits}, we use our
polarization measurements to place a formal upper limit on the
strength of the lensing $B$-mode signal.

Our basic $\Lambda$CDM cosmological model is characterized by the
following six parameters (where $h = H_0 / \left[ 100 \,\, {\rm
kms}^{-1} {\rm Mpc}^{-1} \right]$, $H_0$ being Hubble's constant in
units, kms$^{-1}$Mpc$^{-1}$): the physical baryon density, $\Omega_b
h^2$; the physical cold dark matter density, $\Omega_c h^2$; the ratio
of the sound horizon to the angular diameter distance at last
scattering, $\theta = r_s / D_A$; the optical depth to last
scattering, $\tau$; the scalar spectral index, $n_s$; and the scalar
amplitude, $\mathcal{A}_s = \ln \left[ 10^{10} A_s \right]$. Here,
$A_s$ is the amplitude of the power spectrum of primordial scalar
perturbations, parametrized by $\mathcal{P}_s(k) =
A_s(k/k^s_\star)^{n_s-1}$. We discuss the choice of pivot-point,
$k^s_\star$, in the following section. Other parameters which we
quote, and which are derived from this basic set, are the dark energy
density, $\Omega_\Lambda$ (assumed here to be a simple cosmological
constant), the age of the universe, the total matter density,
$\Omega_m$, the amplitude of matter fluctuations in 8 $h^{-1}$Mpc
spheres, $\sigma_8$, the redshift to reionization, $z_{re}$ and the
value of the present day Hubble constant, $H_0$. For all our
analyses, we assume a flat universe and include the effects of weak
gravitational lensing. We impose the following broad priors on our
base MCMC parameters: $0.005 < \Omega_b h^2 < 0.100$; $0.01 < \Omega_c
h^2 < 0.99$; $0.5 < \theta < 10.0$; $2.7 < \mathcal{A}_s < 4.0$; $0.5
< n_s < 1.5$; $0.01 < \tau < 0.80$. There is also a prior
imposed on the age of the universe ($10 < {\rm Age \left[ Gyrs \right]} < 20$) and on
the Hubble constant ($40 < H_0 \left[ {\rm kms}^{-1} {\rm Mpc}^{-1} \right] < 100$).

We also investigate models extended to include both a running in
the scalar spectral index, $n_{\rm run} = d n_s / d \ln k$ and/or a
possible tensor contribution. Assuming a power law for the tensor
modes, $\mathcal{P}_t \propto k^{n_t}$, we parametrize their amplitude
by the tensor-to-scalar ratio, $r = \mathcal{P}_t / \mathcal{P}_s$. We
adopt a uniform prior measure for $r$ between 0 and 1. For the running
spectral index model, we adopt a prior of $-0.5 < n_{\rm run} <
0.5$ on the running.

\begin{deluxetable*}{ccccccc}
\tabletypesize{\scriptsize}
\tablecaption{ Basic 6 Parameter Constraints } 
\setlength{\tabcolsep}{0.04in} 
\tablehead{ \colhead{} & \colhead{WMAP} & \colhead{WMAP+ACBAR} & \colhead{WMAP+QUaD} & \colhead{WMAP+ACBAR+QUaD} & \colhead{WMAP+ACBAR+QUaD+SDSS}}
\startdata
$        \Omega_bh^2$  & $0.0228^{+0.0006}_{-0.0006}  $&$0.0229^{+0.0006}_{-0.0006}  $&$0.0227^{+0.0005}_{-0.0005} $&$ 0.0227^{+0.0005}_{-0.0005} $&$0.0227^{+0.0005}_{-0.0005}   $    \\
$        \Omega_ch^2$  & $0.109^{+0.006}_{-0.006}     $&$0.111^{+0.006}_{-0.006}     $&$0.108^{+0.006}_{-0.006}    $&$ 0.109^{+0.005}_{-0.005}    $&$0.108^{+0.004}_{-0.004}      $    \\
$          \theta$     & $1.0406^{+0.0030}_{-0.0031}  $&$1.0422^{+0.0027}_{-0.0027}  $&$1.0403^{+0.0024}_{-0.0025} $&$ 1.0415^{+0.0022}_{-0.0023} $&$1.0414^{+0.0022}_{-0.0022}     $    \\
$            \tau$     & $0.090^{+0.017}_{-0.017}     $&$0.090^{+0.017}_{-0.017}     $&$0.090^{+0.017}_{-0.017}    $&$ 0.089^{+0.017}_{-0.017}    $&$0.088^{+0.017}_{-0.016}      $    \\
$             n_s$     & $0.965^{+0.014}_{-0.014}     $&$0.966^{+0.013}_{-0.013}     $&$0.962^{+0.013}_{-0.013}    $&$ 0.962^{+0.013}_{-0.013}    $&$0.962^{+0.012}_{-0.012}      $    \\
$     {\cal A}_s$      & $3.11^{+0.04}_{-0.04}        $&$3.12^{+0.04}_{-0.04}        $&$3.11^{+0.04}_{-0.04}       $&$ 3.11^{+0.04}_{-0.04}       $&$3.11^{+0.03}_{-0.03}         $    \\      
$  \Omega_\Lambda$     & $0.74^{+0.03}_{-0.03}        $&$0.74^{+0.03}_{-0.03}        $&$0.75^{+0.03}_{-0.03}       $&$ 0.75^{+0.03}_{-0.03}       $&$0.75^{+0.02}_{-0.02}         $    \\ 
$             Age$     & $13.68^{+ 0.14}_{- 0.14}     $&$13.63^{+ 0.12}_{- 0.12}     $&$13.69^{+ 0.12}_{- 0.12}    $&$ 13.66^{+ 0.11}_{- 0.11}    $&$13.66^{+ 0.10}_{- 0.10}         $    \\ 
$        \Omega_m$     & $0.26^{+0.03}_{-0.03}        $&$0.26^{+0.03}_{-0.03}        $&$0.25^{+0.03}_{-0.03}       $&$ 0.25^{+0.03}_{-0.03}       $&$0.25^{+0.02}_{-0.02}         $    \\ 
$        \sigma_8$     & $0.80^{+0.04}_{-0.04}        $&$0.80^{+0.03}_{-0.03}        $&$0.79^{+0.03}_{-0.03}       $&$ 0.79^{+0.03}_{-0.03}       $&$0.79^{+0.02}_{-0.02}         $    \\ 
$          z_{re}$     & $10.5^{+ 1.4}_{- 1.4}        $&$10.5^{+ 1.4}_{- 1.3}        $&$10.5^{+ 1.3}_{- 1.3}       $&$ 10.5^{+ 1.3}_{- 1.3}       $&$10.4^{+ 1.4}_{- 1.3}         $    \\ 
$             H_0$     & $72.1^{+ 2.6}_{- 2.6}        $&$72.3^{+ 2.5}_{- 2.5}        $&$72.5^{+ 2.5}_{- 2.5}       $&$ 72.4^{+ 2.4}_{- 2.4}       $&$72.7^{+ 1.7}_{- 1.7}         $    \\   
\enddata
\tablecomments{\small
  We quote the scalar amplitude as $\mathcal{A}_s
  \equiv \ln\left[10^{10} A_s\right]$ for a pivot-point of $k^s_\star = 0.013$ Mpc$^{-1}$.}
\label{tab:params_lcdm}
\end{deluxetable*}

\subsection{Choice of scales (``pivot-points'') for presentation of results}
\label{sec:pivots}
For the primordial power spectrum parametrization which we have
chosen, we need also to choose a scalar pivot-point, $k^s_\star$, the
wavenumber at which $n_s$ and $A_s$ are evaluated. Within standard
$\Lambda$CDM, $n_s$ is modeled as independent of scale and we can map
constraints on $A_s$ obtained at one pivot-point to an arbitrary new
pivot-point, ${k^s_\star}'$, using 
\be 
A_s({k^s_\star}') = A_s(k^s_\star) \,\,\, ({k^s_\star}' /
k^s_\star)^{n_s-1}.  
\ee

For models including a running in the spectral index, both $n_s$ and
$A_s$ are dependent on scale. For these models, we can map
constraints from an old to a new pivot-point using
\ba
A_s({k^s_\star}') &=& A_s(k^s_\star) \,\,\, ({k^s_\star}' /
k^s_\star)^{n_s(k^s_\star)-1 + \frac{1}{2} n_{\rm run}
  \ln({k^s_\star}'/k^s_\star)}, \\
n_s({k^s_\star}') &=& n_s(k^s_\star) + n_{\rm run} \ln({k^s_\star}' / k^s_\star).
\ea

Correlations between the two parameters, $n_s$ and $n_{\rm run}$ are
dependent on the pivot-point at which one chooses to present
results. In particular, there is a scale at which the uncertainties on
these two parameters become uncorrelated \citep{copeland98}. Choosing
to present results at this ``decorrelation scale'' has the attractive
feature that the marginalized 1D constraint on $n_s$ is not degraded
by allowing the running to be non-zero. \cite{finelli06} presented
parameter constraints from CMB and large-scale structure data using a
pivot-point of $k^s_\star = 0.01$~Mpc$^{-1}$ whereas \cite{peiris06}
identified a decorrelation scale of $k^s_\star \approx
0.02$~Mpc$^{-1}$ using the WMAP 3-yr data. 

In order to find the decorrelation pivot-point, we have followed the
analysis of \cite{cortes07} who describe a technique to fit MCMC
chains for the decorrelation scale. They found a decorrelation scale
of $k^s_\star = 0.017$~Mpc$^{-1}$ using the WMAP 3-yr data
set. Repeating their analysis using the WMAP 5-yr data, we find
$k^s_\star = 0.013$~Mpc$^{-1}$. For simplicity, we choose to present
our constraints on $\mathcal{A}_s$ and $n_s$ at this decorrelation
scale for all of the models and dataset combinations which we have
investigated.

For models including a possible tensor component, we still quote our
constraints on $\mathcal{A}_s$ and $n_s$ at $k^s_\star =
0.013$~Mpc$^{-1}$ but for the tensor-to-scalar ratio, $r$, we use a
tensor pivot-point of $k^t_\star = 0.002$~Mpc$^{-1}$. We do not
attempt to remap our constraints on $r$ to a more optimal pivot-point
since the only meaningful data contributing to a constraint on $r$ is
the WMAP temperature power spectrum on very large scales (for which, a
tensor pivot-point of $k^t_\star = 0.002$~Mpc$^{-1}$ is
appropriate). Note also that for these models, we enforce both the
first and second inflation consistency equations
(e.g. \citealt{lidsey97}): $r = - 8 n_t$ and $\frac{d n_t}{d \ln k} =
n_t \left[ n_t - (n_s - 1) \right]$. Additionally enforcing the second
equation ensures that the first consistency equation holds to linear
order in $\Delta \ln k$ on all scales \citep{cortes07}.

\subsection{The concordance $\Lambda$CDM model}
\label{sec:params_lcdm}
\begin{figure}[t]
  \vspace{0.3cm}
  \centering
  \resizebox{0.43\textwidth}{!}{  
    \rotatebox{-90}{\includegraphics{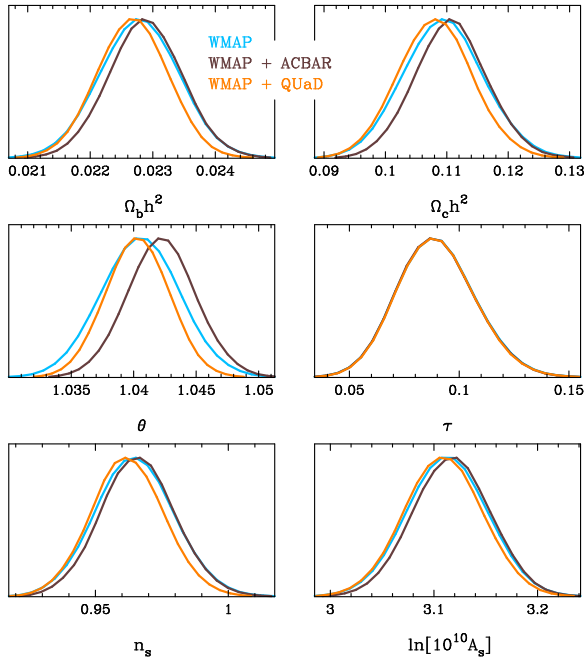}}}
  \caption{1D likelihood distribution for our base MCMC parameters for
  the basic 6-parameter $\Lambda$CDM model. The constraint on the
  scalar amplitude is presented at a pivot-point of $k^s_\star =
  0.013$ Mpc$^{-1}$.}
  \label{fig:params_1d_lcdm}
\end{figure}

In Table~\ref{tab:params_lcdm}, for each dataset combination, we list
the mean recovered values for each parameter, along with their
associated $68\%$ confidence limits, marginalized over all other
parameters. In Figure~\ref{fig:params_1d_lcdm} we plot the
marginalized 1D constraints for the WMAP, WMAP + QUaD and WMAP + ACBAR
combinations. Clearly, the WMAP data dominates when we add in either
the ACBAR or the QUaD data, as was found in our previous analysis
\citep{castro09}. However, the addition of either of these
data sets does provide additional information on $\Omega_b h^2$,
$\Omega_c h^2$ and $\theta$. The biggest improvement in
constraints is in $\theta$ where the WMAP + ACBAR + QUaD combination
tightens the limits by $\sim 25\%$ compared to WMAP alone. This
additional constraining power comes mostly from the QUaD data. 

The mean values of these parameters also shift a little but the only
significant discrepancy is perhaps in the recovered value of
$\theta$. Here, we find the WMAP + ACBAR combination prefers a
somewhat higher value (in agreement with ACBAR's own analysis;
\citealt{reichardt09}) whereas the addition of QUaD data does not
change the WMAP-only preferred mean value but simply tightens the
constraint. 

Note that in comparison to the WMAP team's own analysis
(\citealt{dunkley09}), we recover slightly different mean values for
$\tau$ and more significantly different values for $z_{re}$. This is
due to the different reionization model used in the later version of
the CAMB software which we have used. We note in passing that the
majority of the constraining power in the QUaD data comes from the
measurements of the polarization power spectra as found with our
previous analysis \citep{castro09}.

\subsection{Running spectral index model}
\label{sec:params_run}
\begin{figure*}[t]
  \vspace{0.3cm}
  \centering
  \resizebox{0.9\textwidth}{!}{  
    {\includegraphics{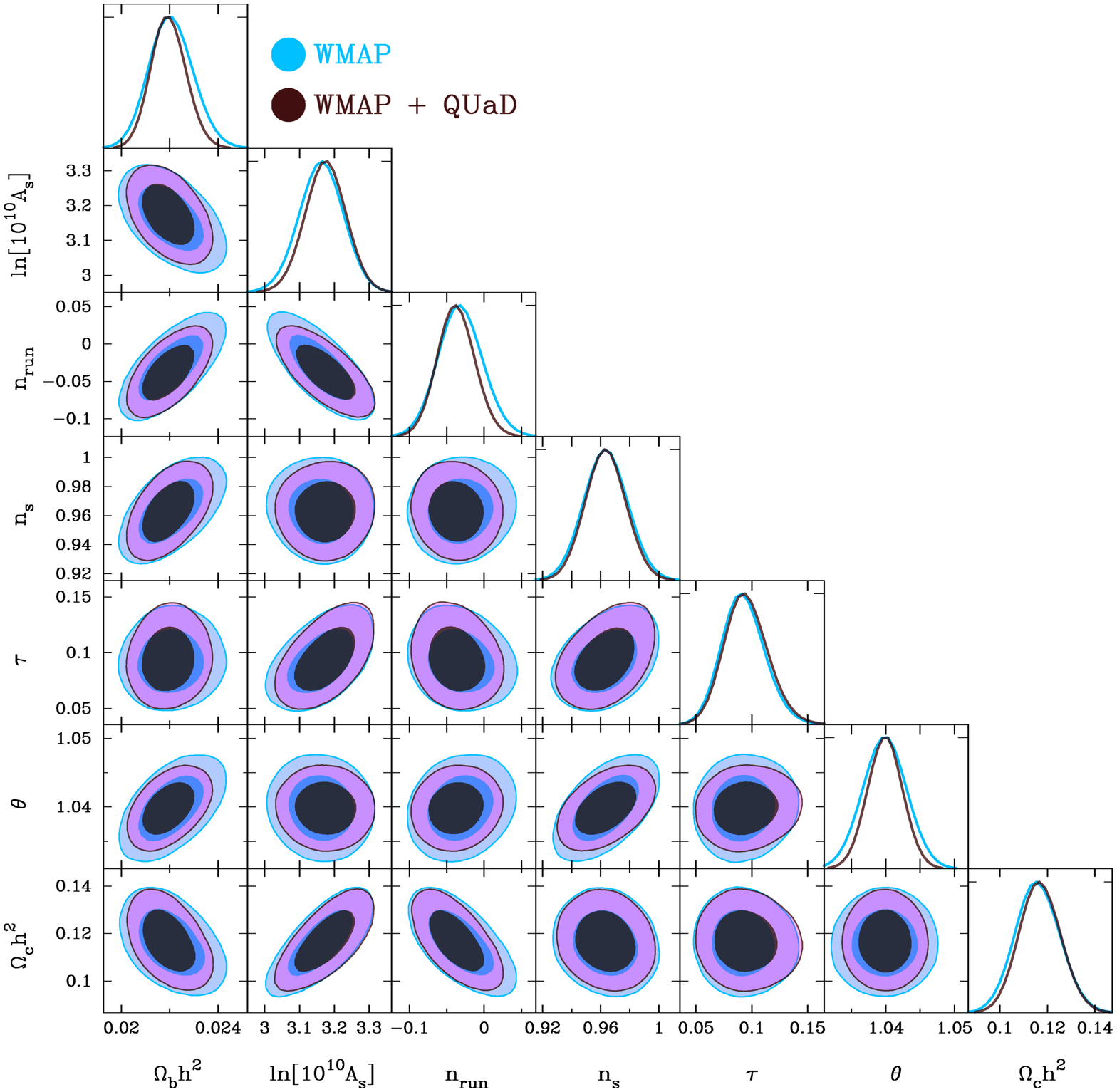}}}
  \caption{1D and 2D likelihood distributions recovered for the base
  MCMC parameters for the $\Lambda$CDM model extended to include a
  possible running in the scalar spectral index. The constraints on
  $n_s$ and $A_s$ are shown for the decorrelation pivot-point of
  $k^s_\star = 0.013$ Mpc$^{-1}$. In the 2D panels, we indicate
  the regions of parameter space which enclose 68\% and 95\% of the
  likelihood as the inner and outer contours respectively. The results
  for the WMAP + QUaD combination are shown over-plotted on the
  WMAP-only results. Adding the QUaD data tightens the constraints on
  $\Omega_b h^2$, $\Omega_c h^2$, $n_{\rm run}$ and $\theta$ by up to
  $20\%$.}
  \label{fig:params_2d_lcdm_run}
\end{figure*}

The 1D and 2D marginalized constraints on our base parameters for the
running spectral index model, as obtained from our WMAP + QUaD runs
are shown in Figure~\ref{fig:params_2d_lcdm_run} along with the
constraints using only the WMAP data. The recovered parameter values
and their uncertainties are listed in
Table~\ref{tab:params_lcdm_run}. 

\begin{figure}[t!]
  \vspace{0.3cm}
  \centering
  \resizebox{0.48\textwidth}{!}{  
    {\includegraphics{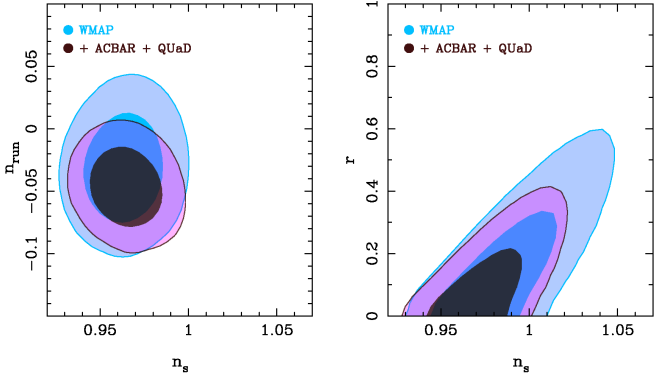}}}
  \caption{Left panel: 68\% and 95\% confidence regions in the
  $n_s$-$n_{\rm run}$ plane, marginalised over all other parameters,
  for the WMAP + ACBAR + QUaD combination as compared to those
  obtained from WMAP alone. The constraints are shown for a
  pivot-point of $k^s_\star = 0.013$ Mpc$^{-1}$. No tensor component
  was allowed for either set of constraints. The constraints tighten
  by about one third. The mean recovered values also shift further
  away from the simple $\{n_s, n_{\rm run}\} = \{1, 0\}$ model. Right
  panel: Marginalized constraints on the inflation parameters, $r$ and
  $n_s$ from WMAP data alone and adding in the ACBAR and QUaD
  datasets. No running in the spectral index was allowed for these
  fits and the tensor-to-scalar ratio, $r$ is presented for a tensor
  pivot-point of $k^t_\star = 0.002$ Mpc$^{-1}$. Once again, the inner
  and outer contours indicate the regions of parameter space enclosing
  68\% and 95\% of the likelhood respectively. The 95\% upper limit on
  $r$ is reduced from $r < 0.48$ to $r < 0.33$. This constraint is
  driven by the preference of the additional datasets for a lower
  spectral index than is recovered from the WMAP data on its own.}
  \label{fig:r_ns_nrun_nomarg}
\end{figure}

The impact of QUaD data is greater for this model --- the QUaD data
adds significantly to the constraints on $\Omega_b h^2$, $\Omega_c
h^2$, $\theta$, and $n_{\rm run}$, reducing the marginalized 1D errors
on these parameters by up to $20\%$. Adding both the QUaD and ACBAR
data has an even greater impact, reducing the errors on these
parameters by up to a third compared to the WMAP-only
uncertainties. Of particular interest are the constraints in the
$n_s$~--~$n_{\rm run}$ plane since many theories of inflation predict
both a deviation from $n_s = 1$ and/or a small negative running.
Constraints from the WMAP + ACBAR + QUaD combination are shown in the
left hand panel of Figure~\ref{fig:r_ns_nrun_nomarg}, together with
the constraints from WMAP alone. Our 1D marginalized constraint on the
running from the combined data set is $n_{\rm run} = 0.046 \pm 0.021$,
$2.2\sigma$ away from the $n_{\rm run} = 0$ model. Adding the LSS data
to the mix improves the constraints even further, in our analysis
tightening the $1\sigma$ error on $n_{\rm run}$ from 0.021 to
0.018. The significance of a non-zero running is also reduced on
addition of the LSS data.

Comparing Tables~\ref{tab:params_lcdm} and \ref{tab:params_lcdm_run},
we see also that the 1D marginalized constraint on the spectral index,
$n_s$ is not weakened by allowing a non-zero running. This is due to
our use of the decorrelation pivot-scale as described in
Section~\ref{sec:pivots}. The results also show that the constraints
on $n_s$ obtained for the standard 6-parameter $\Lambda$CDM model are
robust to marginalization over a possible running. For example, with
the WMAP + ACBAR + QUaD combination, the constraint on $n_s$ goes from
$n_s = 0.962 \pm 0.013$ to $n_s = 0.965 \pm 0.013$ when we allow for a
possible non-zero running. For comparison, if instead, we use the
WMAP-preferred pivot-point ($k^s_\star = 0.002$ Mpc$^{-1}$), the
marginalized 1D constraint on $n_s$ is degraded to $n_s = 0.962 \pm
0.019$ in the presence of running. 

Hints of a negative running in the spectral index have been observed
in previous CMB analyses (e.g. \citealt{dunkley09, reichardt09}). With
the addition of the new QUaD data, this suggestion of a negative
running not only persists, but is strengthened. We do, however,
stress that the combined result shown in the left hand panel of
Fig.~\ref{fig:r_ns_nrun_nomarg} is still consistent with zero running
at the $3\sigma$ level. Nevertheless, it is worth examining the
implications for inflation models if the running was as large and as
negative as the best-fit value returned from our analysis of the
combined CMB data set. In this respect, \cite{malquarti04} have
pointed out that an observed running of $n_{\rm run} \lsim -0.02$
would effectively rule out large field inflation models. More
generally, \cite{easther06} have demonstrated that for a large
negative running, single field slow roll inflation models will last
less than 30 e-folds after entering the horizon. This amount of
inflation is insufficient if inflation happened at the GUT
scale. Consequently, an observation of $n_{\rm run} \sim -0.05$ would
require inflation theory to move beyond the simplest models, e.g. by
considering multiple fields and/or modifications to the slow-roll
formalism (e.g.~\citealt{chung03, makarov05, ballesteros06}).

\subsection{Tensor modes}
\label{sec:params_tens}
Our constraints for the $\Lambda$CDM model including a possible tensor
component are listed in Table~\ref{tab:params_lcdm_tens} in terms of
the mean recovered parameter values and their uncertainties. For $r$,
we quote the 95\% one-tail upper limit since as expected, no detection
of tensors is made. For our WMAP-only analysis, we recover a slightly
weaker limit ($r < 0.48$) than that obtained by the WMAP team
themselves ($r < 0.43$; \citealt{dunkley09}).\footnote{%
Repeating our MCMC analysis using the pre-March 2008 version of CAMB
and adopting WMAP's choice of both scalar and tensor pivot-points, we
recover a result consistent with the WMAP analysis.} Adding either the 
ACBAR or QUaD data, this is reduced to $r < 0.40$. 

The WMAP + ACBAR + QUaD combination produces a constraint on tensor
modes of $r < 0.33$, the strongest from the CMB alone to date. Note
that this constraint does not come from our upper limits on the $BB$
spectrum. It is, in fact, driven by a preference of the small-scale
data (particularly the QUaD $EE$ and $TE$ data) for a somewhat lower
spectral index compared to that preferred by WMAP alone --- a lower
$n_s$ allows more of the large-scale $TT$ power observed by WMAP to
come from scalar perturbations and therefore the maximum allowed
tensor contribution is reduced. Our CMB-only constraints in the
$r$--$n_s$ plane are plotted in the right-hand panel of
Figure~\ref{fig:r_ns_nrun_nomarg}.

\begin{deluxetable*}{ccccccc}
\tabletypesize{\scriptsize}
\tablecaption{ Parameter Constraints for the running spectral index model } 
\setlength{\tabcolsep}{0.04in} 
\tablehead{ \colhead{} & \colhead{WMAP} & \colhead{WMAP+ACBAR} & \colhead{WMAP+QUaD} & \colhead{WMAP+ACBAR+QUaD} & \colhead{WMAP+ACBAR+QUaD+SDSS}}
\startdata
$        \Omega_bh^2$  & $0.0221^{+0.0009}_{-0.0009}  $&$0.0221^{+0.0007}_{-0.0007}   $&$0.0219^{+0.0007}_{-0.0007} $&$ 0.0219^{+0.0006}_{-0.0006} $&$0.0223^{+0.0006}_{-0.0006}   $    \\
$        \Omega_ch^2$  & $0.116^{+0.009}_{-0.009}     $&$0.120^{+0.008}_{-0.008}     $&$0.117^{+0.008}_{-0.008}    $&$ 0.120^{+0.008}_{-0.008}    $&$0.111^{+0.004}_{-0.004}      $    \\
$          \theta$     & $1.0399^{+0.0030}_{-0.0031}  $&$1.0414^{+0.0027}_{-0.0027}  $&$1.0399^{+0.0024}_{-0.0024} $&$ 1.0409^{+0.0023}_{-0.0023} $&$1.0413^{+0.0022}_{-0.0022}      $    \\
$            \tau$     & $0.093^{+0.018}_{-0.018}     $&$0.095^{+0.019}_{-0.019}     $&$0.095^{+0.019}_{-0.018}    $&$ 0.096^{+0.019}_{-0.019}    $&$0.096^{+0.019}_{-0.018}      $    \\
$             n_s$     & $0.964^{+0.014}_{-0.014}     $&$0.967^{+0.014}_{-0.014}     $&$0.963^{+0.013}_{-0.013}    $&$ 0.965^{+0.013}_{-0.013}    $&$0.967^{+0.013}_{-0.013}      $    \\
$    n_{\rm run}$     & $-0.031^{+0.028}_{-0.028}    $&$-0.040^{+0.023}_{-0.023}    $&$-0.038^{+0.024}_{-0.024}   $&$ -0.046^{+0.021}_{-0.021}   $&$-0.028^{+0.018}_{-0.018}      $    \\
$     {\cal A}_s$      & $3.16^{+0.06}_{-0.06}        $&$3.18^{+0.05}_{-0.05}        $&$3.17^{+0.06}_{-0.06}       $&$ 3.19^{+0.05}_{-0.05}       $&$3.15^{+0.04}_{-0.04}         $    \\      
$  \Omega_\Lambda$     & $0.70^{+0.05}_{-0.05}        $&$0.69^{+0.05}_{-0.05}        $&$0.70^{+0.05}_{-0.05}       $&$ 0.69^{+0.05}_{-0.05}       $&$0.74^{+0.02}_{-0.02}         $    \\ 
$             Age$     & $13.81^{+ 0.18}_{- 0.18}     $&$13.78^{+ 0.14}_{- 0.15}     $&$13.82^{+ 0.14}_{- 0.14}    $&$ 13.82^{+ 0.13}_{- 0.13}    $&$13.72^{+ 0.11}_{- 0.11}         $    \\ 
$        \Omega_m$     & $0.30^{+0.05}_{-0.05}        $&$0.31^{+0.05}_{-0.05}        $&$0.30^{+0.05}_{-0.05}       $&$ 0.31^{+0.05}_{-0.05}       $&$0.26^{+0.02}_{-0.02}         $    \\ 
$        \sigma_8$     & $0.81^{+0.04}_{-0.04}        $&$0.83^{+0.03}_{-0.03}        $&$0.81^{+0.04}_{-0.04}       $&$ 0.83^{+0.03}_{-0.03}       $&$0.79^{+0.02}_{-0.02}         $    \\ 
$          z_{re}$     & $11.2^{+ 1.6}_{- 1.6}        $&$11.4^{+ 1.6}_{- 1.6}        $&$11.4^{+ 1.6}_{- 1.6}       $&$ 11.7^{+ 1.6}_{- 1.6}       $&$11.2^{+ 1.5}_{- 1.5}         $    \\ 
$             H_0$     & $68.9^{+ 4.0}_{- 4.0}        $&$68.0^{+ 3.4}_{- 3.4}        $&$68.5^{+ 3.5}_{- 3.6}       $&$ 67.5^{+ 3.2}_{- 3.2}       $&$71.4^{+ 1.9}_{- 1.9}         $    \\   
\enddata
\tablecomments{\small
  The pivot point used for $\mathcal{A}_s$ and $n_s$ is $k^s_\star =
  0.013$ Mpc$^{-1}$.}
\label{tab:params_lcdm_run}
\end{deluxetable*}

\subsection{Running spectral index and tensor modes}
\label{sec:params_run_tens}
\begin{figure}[t!]
  \vspace{0.3cm}
  \centering
  \resizebox{0.48\textwidth}{!}{  
    {\includegraphics{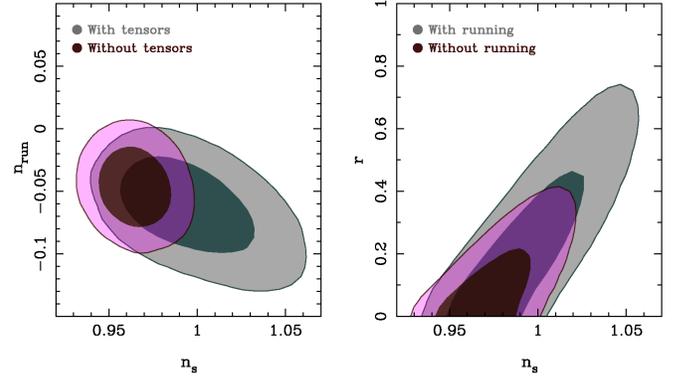}}}
  \caption{Left panel: Constraints are shown (as 68\% and 95\%
  confidence regions) in the $n_{\rm run}$--$n_s$ plane with and
  without marginalization over a possible tensor component. These fits
  are for the WMAP + ACBAR + QUaD combination. Allowing a non-zero
  tensor component weakens the constraints considerably. However, the
  addition of both QUaD and ACBAR to WMAP still favors a small
  negative running. Right panel: The 68\% and 95\% central confidence
  regions in the $r$--$n_s$ plane (for WMAP + ACBAR + QUaD) with and
  without marginalization over a possible running in the spectral
  index. Allowing the spectral index to run degrades the constraints
  to such an extent that the addition of QUaD and/or ACBAR data yields
  no improvement over the WMAP-only constraints. $n_s$ is evaluated at
  $k^s_\star = 0.013$ Mpc$^{-1}$ (for both panels) and $r$ is
  evaluated at $k^t_\star = 0.002$ Mpc$^{-1}$.}
  \label{fig:r_ns_nrun_marg}
\end{figure}

When we allow for both a running in the spectral index \emph{and} a
tensor contribution the constraints weaken considerably versus either
on their own. In the left-hand panel of
Figure~\ref{fig:r_ns_nrun_marg}, for the WMAP + ACBAR + QUaD
combination, we plot constraints in the $n_{\rm run}$--$n_s$ plane
with and without marginalization over a possible tensor component. The
right panel of this figure shows the corresponding constraints in the
$r$--$n_s$ plane with and without marginalization over a possible
running in the spectral index. For this model, the addition of QUaD
and ACBAR data still improves the constraints in the spectral index
running (and indeed, still strongly suggests a small negative running)
but the constraints in the $r$--$n_s$ plane in the presence of running
do not improve on the WMAP-only result. This degradation in the
constraints on $r$ when we allow for a running in the spectral index
is further demonstrated in Table~\ref{tab:params_run_tens_small} where
we quote the 1D marginalized constraints on the parameters $\{r, n_s$,
$n_{\rm run}\}$ for the tensors-only, running-only and tensors +
running models. In this table, we present the results for the WMAP +
ACBAR + QUaD combination and for the case where we add in the SDSS LRG
data.

\subsection{Constraints on parity violation}
\label{sec:params_parviol}
In the preceding sections, we used the QUaD $TT, EE, TE$ and $BB$
spectra to constrain the parameters of standard $\Lambda$CDM models
and its usual extensions. For that analysis, we did not use our
measurements of the cross-polarization spectrum ($EB$) or the
correlation of temperature with $B$-modes ($TB$), since these spectra
are expected to vanish in a universe which respects parity
conservation (which the above models do). In this section, we use
these two spectra (along with the $TE$, $EE$ and $BB$ measurements) to
constrain a possible parity violation signal on cosmological
scales. In the presence of parity violating interactions, a rotation
in the polarization direction of CMB photons will be induced as they
propagate from the surface of last scattering. If parity violating
effects are present on cosmological scales, there will therefore be a
net local rotation of the observed Stokes parameters, $Q$ and $U$ in
the measured polarization map. This will mix $E$ and $B$
modes resulting in non-zero expectations for the $TB$ and $EB$ power
spectra. Parametrizing the parity violation effect with a rotation
angle, $\Delta \alpha$, the expectation values for the $TB$ and $EB$
spectra in terms of the cosmological $TE$ and $EE$ spectra are given by:
\ba
C_\ell^{TB} &=& C_\ell^{TE} \sin \, (2 \Delta \alpha) \\
C_\ell^{EB} &=& \frac{1}{2} C_\ell^{EE} \sin \, (4 \Delta \alpha) \, .
\ea
In addition, assuming that primordial and lensed $B$-modes are
negligible (which is an excellent assumption given our sensitivity), 
the expectation value for the $BB$ spectrum is 
\be
C_\ell^{BB} = C_\ell^{EE} \sin^2(2 \Delta \alpha) \, .
\ee
We used our previous results to place a constraint of $\Delta
\alpha = 0.53^\circ \pm 0.82^\circ \pm 0.50^\circ$ \citep{wu09} where
the two quoted uncertainties are the random and systematic components
respectively.\footnote{%
The parity violation effect is completely degenerate with an error in
the calibration of the polarization co-ordinate system of the
experiment. As described in \cite{wu09}, we are confident in our
calibration to at least $0.5^\circ$.} 

Here, we repeat this analysis with our new measurements.
The analysis is model-independent in the sense that
we construct our estimator for $\Delta \alpha$ in terms of the
observed power spectra and do not assume a cosmological model for the
$EE$ or $TE$ signals. For details of our estimator and analysis (which
has not changed since our previous work), we refer the reader to
\cite{wu09}.

\begin{figure}[t]
  \vspace{0.3cm}
  \centering
  \resizebox{0.43\textwidth}{!}{  
    \rotatebox{-90}{\includegraphics{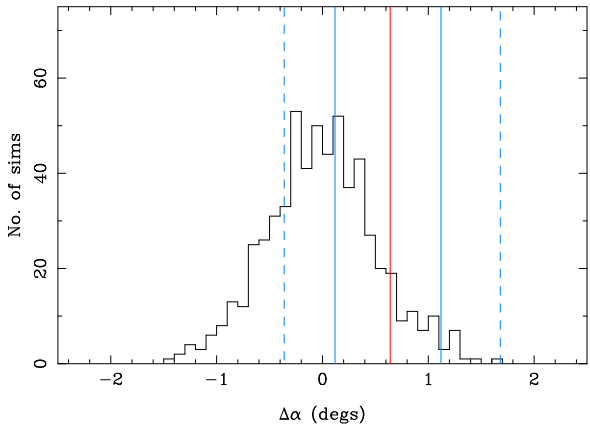}}}
  \caption{Constraints on possible parity violation interactions on
  cosmological scales, parametrized in terms of the parity violation
  rotation angle, $\Delta \alpha$. The histogram shows the estimates
  of $\Delta \alpha$ as measured from our suite of signal + noise 
  simulations. The vertical red line shows $\Delta \alpha$ as measured
  from the real QUaD data. The full and dashed blue lines show the
  68\% and 95\% confidence regions about the central value as
  estimated from the scatter in the simulation results.}
  \label{fig:parviol}
\end{figure}

We apply the estimator to the real QUaD data and assign error-bars due
to random noise and sample variance by processing the suite of
simulations containing both signal and noise through the
analysis. The result is shown in Figure~\ref{fig:parviol} where we
plot both the results from simulations and from the real data. Note
that our simulations contain no parity violating signals and so should
scatter about zero, which they do. We take the scatter in the results
from the simulations as our random error. Adding in the systematic
error, our final result is 
\be
\Delta \alpha = 0.64 \pm 0.50 \mbox{ (random)} \pm 0.50
\mbox{ (systematic)} \, .
\ee
The random error has been reduced by $\sim 40\%$ with respect to our
previous analysis in line with expectations. 

Our result can be compared to the limits obtained from the WMAP 5-year
data ($\Delta \alpha = -1.7 \pm 2.1$; \citealt{komatsu09}) or to the
limits obtained from the combination of the WMAP 5-year data and the
B03 results ($\Delta \alpha = -2.6 \pm 1.9$; \citealt{xia08}). We note
that both of these quoted results include random errors only and do
not include estimates of the systematic errors on the WMAP and B03
polarization calibration angles. Even when we include this systematic
uncertainty for QUaD, our result is clearly a marked improvement over
these previous analyses.

\begin{deluxetable*}{ccccccc}
\tabletypesize{\scriptsize}
\tablecaption{ Parameter constraints including a possible tensor component } 
\setlength{\tabcolsep}{0.04in} 
\tablehead{ \colhead{} & \colhead{WMAP} & \colhead{WMAP+ACBAR} & \colhead{WMAP+QUaD} & \colhead{WMAP+ACBAR+QUaD} & \colhead{WMAP+ACBAR+QUaD+SDSS}}
\startdata
$        \Omega_bh^2$  & $0.0235^{+0.0008}_{-0.0008}  $&$0.0234^{+0.0007}_{-0.0007}  $&$0.0232^{+0.0007}_{-0.0007} $&$ 0.0231^{+0.0006}_{-0.0006} $&$0.0229^{+0.0006}_{-0.0006}   $    \\
$        \Omega_ch^2$  & $0.104^{+0.007}_{-0.007}     $&$0.106^{+0.007}_{-0.007}     $&$0.103^{+0.007}_{-0.007}    $&$ 0.105^{+0.006}_{-0.006}    $&$0.107^{+0.004}_{-0.004}      $    \\
$          \theta$     & $1.0421^{+0.0033}_{-0.0033}  $&$1.0433^{+0.0028}_{-0.0028}  $&$1.0412^{+0.0026}_{-0.0026} $&$ 1.0423^{+0.0023}_{-0.0023} $&$1.0419^{+0.0022}_{-0.0023}      $    \\
$            \tau$     & $0.094^{+0.018}_{-0.018}     $&$0.092^{+0.018}_{-0.018}     $&$0.093^{+0.018}_{-0.018}    $&$ 0.091^{+0.017}_{-0.017}    $&$0.088^{+0.017}_{-0.017}      $    \\
$             n_s$     & $0.990^{+0.023}_{-0.023}     $&$0.986^{+0.021}_{-0.020}     $&$0.982^{+0.020}_{-0.020}    $&$ 0.978^{+0.018}_{-0.018}    $&$0.973^{+0.015}_{-0.015}      $    \\
$     {\cal A}_s$      & $3.07^{+0.05}_{-0.05}        $&$3.08^{+0.04}_{-0.04}        $&$3.08^{+0.04}_{-0.04}       $&$ 3.09^{+0.04}_{-0.04}       $&$3.09^{+0.03}_{-0.04}         $    \\      
$              r$      & $<0.48$ (95\% c.l.)           &$<0.40$ (95\% c.l.)           &$<0.40$ (95\% c.l.)          &$<0.33$ (95\% c.l.)           &$<0.27$ (95\% c.l.)   \\      
$  \Omega_\Lambda$     & $0.78^{+0.03}_{-0.03}        $&$0.77^{+0.03}_{-0.03}        $&$0.78^{+0.03}_{-0.03}       $&$ 0.77^{+0.03}_{-0.03}       $&$0.76^{+0.02}_{-0.02}         $    \\ 
$             Age$     & $13.52^{+ 0.18}_{- 0.18}     $&$13.51^{+ 0.16}_{- 0.16}     $&$13.57^{+ 0.15}_{- 0.15}    $&$ 13.57^{+ 0.13}_{- 0.13}    $&$13.61^{+ 0.11}_{- 0.11}         $    \\ 
$        \Omega_m$     & $0.22^{+0.03}_{-0.03}        $&$0.23^{+0.03}_{-0.03}        $&$0.22^{+0.03}_{-0.03}       $&$ 0.23^{+0.03}_{-0.03}       $&$0.24^{+0.02}_{-0.02}         $    \\ 
$        \sigma_8$     & $0.77^{+0.04}_{-0.04}        $&$0.78^{+0.04}_{-0.04}        $&$0.76^{+0.04}_{-0.04}       $&$ 0.78^{+0.03}_{-0.03}       $&$0.78^{+0.02}_{-0.02}         $    \\ 
$          z_{re}$     & $10.5^{+ 1.4}_{- 1.4}        $&$10.5^{+ 1.4}_{- 1.4}        $&$10.5^{+ 1.4}_{- 1.3}       $&$ 10.4^{+ 1.4}_{- 1.3}       $&$10.3^{+ 1.3}_{- 1.3}         $    \\ 
$             H_0$     & $75.8^{+ 3.8}_{- 3.8}        $&$75.2^{+ 3.4}_{- 3.4}        $&$75.5^{+ 3.4}_{- 3.4}       $&$ 74.8^{+ 3.1}_{- 3.1}       $&$73.8^{+ 1.8}_{- 1.9}         $    \\   
\enddata
\tablecomments{\small
  The pivot point used for $\mathcal{A}_s$ is $k^s_\star =
  0.013$ Mpc$^{-1}$ while the pivot point used for the tensor-to-scalar
  ratio, $r$ is $k^t_\star = 0.002$ Mpc$^{-1}$.}
\label{tab:params_lcdm_tens}
\end{deluxetable*}

\begin{deluxetable}{c c c c}
\tablecaption{Constraints on inflationary parameters}
\tablehead{\colhead{Parameter} & \colhead{Tensors} & \colhead{Running} & \colhead{Tensors + Running}}
\startdata
CMB only: &                               &                    &                            \\
$r$                  & < 0.33 (95\% c.l.) &                    &  < 0.60 (95\% c.l.)        \\
$d n_s / d \ln k$    &                    & $-0.046^{+0.021}_{-0.021}$ &  $-0.063^{+0.025}_{-0.025}$        \\
$n_s$                & $0.978^{+0.018}_{-0.018}$  & $0.965^{+0.013}_{-0.013}$  &  $0.997^{+0.026}_{-0.025}$ \\
                     &                    &                    &		     \\
\hline
                     &                    &                    &		     \\
CMB + LSS: &                              &                    &                            \\
$r$                  & < 0.27 (95\% c.l.) &                    &  < 0.61 (95\% c.l.)        \\
$d n_s / d \ln k$    &                    & $-0.028^{+0.018}_{-0.018}$ &  $-0.052^{+0.023}_{-0.023}$        \\
$n_s$                & $0.973^{+0.015}_{-0.015}$  & $0.967^{+0.013}_{-0.013}$  &  $0.999^{+0.024}_{-0.024}$         \\

\enddata
\label{tab:params_run_tens_small}
\end{deluxetable}

\subsection{Limits on the lensed $B$-mode signal}
\label{sec:params_bmode_limits}
Although QUaD has not made any detection of $B$-modes, it is the most
sensitive small-scale CMB polarization experiment to date. We can
therefore place the leading upper limit on the presence of a small
scale $B$-mode signal. The signal expected to dominate on the scales
at which QUaD is sensitive is that induced by gravitational lensing of
$E$-modes by intervening large scale structure. As well as measuring
cosmological $B$-modes from inflation, future polarization experiments
will target this lensing signal from which useful information can be
gained on dark energy and massive neutrinos
(e.g. \citealt{kaplinghat03, smith06}).

Assuming a single flat band power between $\ell = 200$ and $\ell =
2000$, we find $\ell(\ell + 1)C^{BB}_\ell/2\pi = 0.17 \pm 0.17 \, \mu {\rm
K}^2$ with a 95\% upper limit of $0.57 \, \mu {\rm K}^2$. The errors 
quoted are estimated from the scatter in the
results obtained from the suite of simulations containing both
signal and noise. For comparison, the $\Lambda$CDM expectation value
for this band power is $0.058 \, \mu {\rm K}^2$. Alternatively,
assuming the $\Lambda$CDM shape for the lensing signal, and simply
fitting for its amplitude between $\ell = 200$ and $\ell = 2000$, our
constraint on the amplitude\footnote{%
Our normalization convention is such that the amplitude of the lensed
$B$-mode signal in the concordance $\Lambda$CDM model is unity.}
is $2.5 \pm 4.5$ with a 95\% upper limit of $12.5$.

Note that although we have used all of our $BB$ band powers to obtain
the above constraints, the window function of our estimator is
strongly skewed towards lower multipoles where the band power uncertainties are much
smaller. The effective range of multipoles to which our upper limits
apply is, in fact, $170 < \ell < 400$ rather than the nominal $200 <
\ell < 2000$ range. 

Although our $2\sigma$ upper limits are an order of magnitude larger
than the expected $\Lambda$CDM signal, they are, in turn, roughly an
order of magnitude better than previously reported limits on the
amplitude of the $B$-mode signal in this angular scale range. 

\section{Conclusions}
\label{sec:conclusions}
We have presented a re-analysis of the final dataset from the QUaD
experiment, a CMB polarimeter which observed the CMB at 100 and
150~GHz from the South Pole between 2005 and 2007. A major part of this 
re-analysis was the development of a new technique for removing ground
contamination from the data. The ground signal seen in QUaD data
is polarized and, if not removed, contaminates all of the CMB
power spectrum measurements. Our new procedure, which is based on
constructing, and subsequently subtracting, templates of the ground
signal has allowed us to reconstruct maps of the $T$, $Q$ and $U$
Stokes parameters over the full sky area. Although the method is
not entirely lossless, it provides, on average, a 30\% increase in the
precision of the power spectra compared to our previous analysis
which used field-differencing to remove the ground. 

Through further detailed analysis of calibration data, we have also
significantly improved our understanding of the QUaD beams. We have
implemented new beam models which explicitly incorporate the effects
of sidelobes, resulting in an increase of $\sim 10\%$ in the amplitude of our
power spectra measurements for multipoles, $\ell \gsim 700$. The shift
in power is most relevant for our high-$\ell$ temperature power
spectrum measurements where the signal-to-noise is high.

We have presented results using our two independent analysis
pipelines. Though there are significant differences in the approach
between the two pipelines, the final results agree very well. Testing
the power spectra against the best-fit $\Lambda$CDM model to the WMAP
5-year data, we find good agreement. Our measurements of the
$E$-mode polarization spectrum, and of the cross-correlation between
the $E$-modes and the CMB temperature field, are the most precise at
multipoles $\ell > 200$ to date. Our measurement of the temperature
power spectrum at $\ell > 1000$ is among the best constraints on
temperature anisotropies on small angular scales and is competitive
with the final ACBAR result \citep{reichardt09}.

We have subjected our results to the same set of rigorous jackknife
tests for systematic effects as was performed in our previous analysis
\citep{pryke09}. We find no evidence for residual systematic effects
in our polarization maps. Although formally, many of our $TT$
jackknife tests fail, the inferred levels of residual systematics are
negligible compared to our sample-variance driven error
bars. Moreover, the very small level of power seen in our frequency
difference maps and power spectra indicate that foreground
contamination is also negligible compared to our uncertainties.

We have used our power spectra measurements, in combination with the
WMAP 5-year results and the ACBAR results to place
constraints on the parameters of cosmological models. For the standard
6-parameter $\Lambda$CDM model, the QUaD data adds only marginally to
the constraints obtained from the WMAP data alone. The impact of the
QUaD data is greater in a model extended to include a running in the
spectral index, reducing the uncertainties in $\Omega_b h^2$, $\Omega_c h^2$, $\theta$ and $n_{\rm
run}$ by up to 20\%. The addition of both
QUaD and ACBAR data is more powerful still, improving the 
constraints on these four parameters by up to one third. For a
$\Lambda$CDM model extended to include a possible tensor component, we
find that the addition of both ACBAR and QUaD data reduces the upper
limit on the tensor-to-scalar ratio from $r < 0.48$ to $r < 0.33$
(95\% c.l.). This is the strongest limit to date on tensors from the
CMB alone. The improvement is driven by a tendency of the QUaD data to
prefer a somewhat smaller spectral index than is inferred from WMAP
data alone. 

We have used our measurements of the $TB$ and $EB$
power spectra to put constraints on possible parity-violating
interactions on cosmological scales. Following our previous analysis 
\citep{wu09}, we constrain the rotation angle due to such a possible 
``cosmological birefringence'' to be 
$0.64^\circ \pm 0.50^\circ \pm 0.50^\circ$ where the errors quoted are
the random and systematic contributions. Our result is equivalent to a
constraint on isotropic Lorentz-violating interactions of
$k^{(3)}_{(V)00} < 1.5 \times 10^{-43}$ GeV (68\% c.l.).
 
Finally, we have placed an upper limit on the strength of the
lensing $B$-mode signal using our measurements of the $BB$ power
spectrum. Assuming the concordance $\Lambda$CDM shape for lensing
$B$-modes, we constrain its amplitude (where the normalization is such
that the $\Lambda$CDM model has amplitude $=1$) to be $2.5 \pm 4.5$ with a 95\%
upper limit of 12.5. Alternatively, assuming a single flat band power
for $\ell > 200$ we find a 95\% upper limit of
$\ell(\ell + 1) C^{BB}_\ell / 2 \pi < 0.57 \, \mu{\rm K}^2 $ for the amplitude of
$B$-modes.

\acknowledgements QUaD is funded by the National Science Foundation in
the USA, through grants ANT-0338138, ANT-0338335 \& ANT-0338238, by
the Science and Technology Facilities Council (STFC) in the UK and by
the Science Foundation Ireland. The BOOMERanG collaboration kindly
allowed the use of their CMB maps for our calibration purposes. PGC
acknowledges funding from the Portuguese FCT. SEC acknowledges support
from a Stanford Terman Fellowship. JRH acknowledges the support of an
NSF Graduate Research Fellowship, a Stanford Graduate Fellowship and a
NASA Postdoctoral Fellowship. YM acknowledges support from a SUPA
Prize studentship. CP acknowledges partial support from the Kavli
Institute for Cosmological Physics through the grant NSF PHY-0114422.
EYW acknowledges receipt of an NDSEG fellowship. MZ acknowledges
support from a NASA Postdoctoral Fellowship. Part of the analysis
described in this paper was carried out on the University of
Cambridge's distributed computing facility, CAMGRID. We acknowledge
the use of the FFTW \citep{frigo05}, CAMB \citep{lewis00}, CosmoMC
\citep{lewis02} and HEALPix \citep{gorski05} packages. We acknowledge
the use of the Legacy Archive for Microwave Background Data Analysis
(LAMBDA). Support for LAMBDA is provided by the NASA Office of Space
Science.

\begin{appendix}

\section{Beam uncertainties} 
\label{app:beam_model_app}
As described in Section~\ref{sec:beam_model}, our new beam models
involve either fitting the QUaD physical optics (PO) beam models to
QSO data (Pipeline 1) or measuring the sidelobes directly from 
QSO maps under the assumption that the sidelobes are azimuthally
symmetric (Pipeline 2). Although the predicted radially averaged beam
profiles from both of these approaches appear to match the data
very well, the fits are not perfect and are subject to an uncertainty
in the sidelobe levels. There is also an uncertainty on the width of
the main lobe, dominated by small temperature-dependent variations. Based on
the fluctuations in the beam widths seen in our ``rowcal''
data\footnote{%
These calibration data consisted of scanning each row of pixels
in the focal plane across the bright HII region, RCW38 and were taken
daily throughout the QUaD observations. Although RCW38 is not a true point
source, the fluctuations in the per-channel beam widths put a tight
constraint on temperature dependent seasonal fluctuations in our main lobe beams.}, 
we estimate the remaining uncertainty on the main lobe width to
be 2.5\% of the effective FWHMs of 5.2 and 3.8 arcmin at 100~GHz
and 150~GHz respectively. We obtain the uncertainty on the level of our
sidelobes from the errors returned from fitting our PO simulations to
the QSO observations in Pipeline 1.

\begin{figure}[t]
  \vspace{0.3cm}
  \centering
  \resizebox{0.50\textwidth}{!}{  
    \rotatebox{-90}{\includegraphics{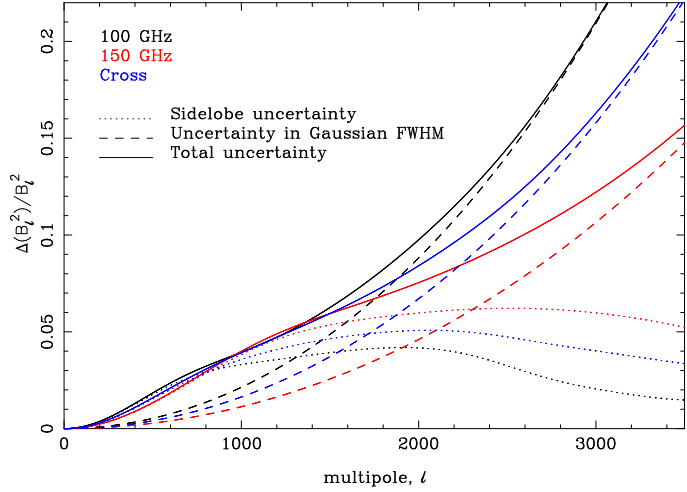}}}
  \caption{Fractional uncertainties on our beam transfer
  functions. The dashed curves show the errors due to the 2.5\% uncertainty
  in the width of the main lobe beams. The dotted curves show the
  sidelobe uncertainties and the full curves show the total
  uncertainties. The corresponding curves for the combined spectra are
  similar to the 150~GHz curves shown in red. At 150~GHz, the
  uncertainty in the level of the sidelobes dominates the beam
  uncertainty for the full $\ell$-range presented in this paper.}
  \label{fig:beam_uncert}
\end{figure}

To propagate these errors onto uncertainties in the transfer
functions of Section~\ref{sec:transfer}, for the error in the main
lobe, we simply note that the effect of a fractional error, $\delta$
in the FWHM of a Gaussian beam is well approximated by 
\be
\frac{\Delta B^2_\ell}{B^2_\ell} \, = \, \exp \left[ \sigma^2_b (\delta^2 + 2
\delta) \ell (\ell + 1) \right] \, - \, 1 \, ,
\label{eqn:beam_err1}
\ee 
where $\sigma_b = \theta_{\rm FWHM} / \sqrt{8 \ln 2}$ is the beam
width. For the errors in the sidelobes, we take the minimum and
maximum sidelobe levels as returned from the fits of the PO models to
the data, coadd the resulting beam models across detectors and
radially average to produce the minimum and maximum allowed radial
profiles, $B(\theta)$, for each frequency. Taking the Legendre transform
of these profiles,
\be
B_\ell = 2 \pi \int B(\theta) P_\ell \cos(\theta) \, d \! \cos(\theta)
\, ,
\ee 
we estimate the error in our beam transfer functions due to the
uncertainty in the sidelobes as
\be
\Delta B^{2}_\ell = B_{{\rm max}, \ell}^2 - B_{{\rm min}, \ell}^2 \, .
\label{eqn:beam_err2}
\ee 
We take the quadrature sum of the errors due to the main lobe and
sidelobe uncertainties to be the final error. These uncertainties are
shown in Figure~\ref{fig:beam_uncert} along with the quadrature
sum. Since our combined spectra are dominated by the 150~GHz channel,
the curves for the combined spectra are approximately the same as the
150~GHz curves.

\section{Absolute calibration uncertainty} 
\label{app:abs_cal_app}
As described in Section~\ref{sec:abs_cal}, our calibration is
performed by taking the ratio of cross-spectra between the QUaD and
B03 temperature maps. This process is subject to several
uncertainties.

Firstly, there is a statistical error in the calibration ratio,
predominantly due to noise in the B03 maps. To estimate this error, we
perform Monte-Carlo simulations of the absolute calibration
process. Assuming white noise and a $\Lambda$CDM power spectrum, we
use the B03 hit-maps along with their stated sensitivities to produce
simulations of the B03 maps. We do the same thing for QUaD and apply a
known $\mu$K $\rightarrow$ V calibration to the simulated QUaD
maps. Each pair of simulated maps is then passed through the absolute
calibration analysis. The scatter in the calibration factors recovered
from these simulations is 0.6\% and we take this as the statistical
uncertainty in our calibration.

Our calibration ratio as a function of multipole is not perfectly flat
but fluctuates about a mean value. Although some of this scatter
will be due to noise (which is included in our estimate of the
statistical error), we conservatively also include this scatter,
which we measure to be 1.1\%, in our error budget. In addition, we
have also performed the calibration analysis using each of the
jackknife splits described in Section~\ref{sec:jackknives}. Although
the scatter found in the recovered calibration numbers do not indicate
any significant inconsistencies, we also include this scatter in our error
budget.  

To propagate the errors in the B03 and QUaD beam transfer functions
onto our calibration, we repeat the analysis but with the
beam functions shifted by their quoted errors. Doing this for each of
the B03 and QUaD beams, we take the resulting shifts in the
calibration numbers as the error due to uncertainty in the beams. 
We find a 1.1\% shift due to the uncertainty in the B03 beam function
and a 0.75\% shift due to the uncertainty in QUaD's beam.

A further source of error is the relative pointing uncertainty between
the QUaD and B03 maps. There is a clear pointing offset seen between the QUaD
and B03 maps and so we have shifted the B03 maps before performing the
calibration analysis. We find the appropriate shift (which we model as
a simple shift in R.A. and Dec) by fitting for it in map-space. To
quantify the error, we repeat the analysis with the B03 maps shifted
(away from the best fit) according to the errors returned from our
map-based fit. Applying the shift in a number of different directions
(the eight compass points), we find the maximum shift in the
resulting calibration factor is 1.6\%. We take this number as our
error due to the B03/QUaD relative pointing uncertainty.

Finally, QUaD also inherits the stated uncertainty in the B03 
calibration which is 2\% (\citealt{masi06}). We add this and each of
the errors derived above in quadrature to arrive at our
final calibration uncertainty of 3.4\%. Our absolute calibration
error budget is summarized in Table~\ref{tab:abs_cal}.

\begin{deluxetable}{c c}
\tablecaption{Absolute calibration uncertainties for QUaD}
\tablehead{\colhead{Source} & \colhead{Uncertainty(\%)} }
\tablewidth{10cm}
\startdata
Statistical error in calibration ratio   & 0.60 \\
$\ell$-dependence of calibration ratio   & 1.10 \\
Uncertainty in B03 $B_\ell$              & 1.10 \\
Uncertainty in QUaD $B_\ell$             & 0.75 \\
Pointing uncertainty                     & 1.62 \\
Internal consistency (jackknifes)        & 1.20 \\
B03 calibration error                    & 2.00 \\
                                                \\
Total uncertainty                        & 3.38 \\

\enddata
\label{tab:abs_cal}
\end{deluxetable}

\end{appendix}

\bibliographystyle{apj}
\bibliography{ms}
\end{document}